# Multi-Fidelity Predictive Model for Shock Response of Energetic Materials Using Conditional U-Net

Brian H. Lee,[1] Chunyu Li,[1] Aidan Pantoya,[1] James P. Larentzos,[2] John K. Brennan,[2] Alejandro Strachan[1, a)]

Author affiliations
[1]*School of Materials Engineering and Birck Nanotechnology Center, Purdue University, West Lafayette, Indiana 47907, USA*
[2]*U.S. Army Combat Capabilities Development Command (DEVCOM) Army Research Laboratory, Aberdeen Proving Ground, Maryland 21005, USA*

Author email
[a)]Author to whom correspondence should be addressed: strachan@purdue.edu

## Abstract

Mapping microstructure to properties is central to materials science. Perhaps most famously, the Hall-Petch relationship relates average grain size to strength. More challenging has been deriving relationships for properties that depend on subtle microstructural features and not average properties. One such example is the initiation of energetic materials under dynamical loading, dominated by energy localization on microstructural features such as pores, cracks, and interfaces. We propose a conditional convolutional neural network to predict the shock-induced temperature field as a function of shock strength, for a wide range of microstructures, and obtained via two different simulation methods. The proposed model, denoted MISTnet2, significantly extends prior work that was limited to a single shock strength, model, and type of microstructure. MISTnet2 can contribute to bridging atomistics with coarse-grain simulations and enable first principles predictions of detonation initiation and safety of this class of materials.

# 1. Introduction

The initiation of polymer-bonded explosives (PBXs) under shock loading is a complex, multi-scale phenomenon governed by the material's microstructure.[1, 2] The shock-to-deflagration transition of PBXs is driven by the formation of hotspots, where mechanical energy localizes from the interaction of the shock and microstructural features. Various mechanisms with relevant length scales ranging from a few nm to microns— including the collapse of pores,[3, 4] interfacial friction,[5] crack propagation,[6, 7] plasticity,[8, 9] and the interaction thereof[10]— affect the shape, temperature, size and interaction of the hotspots, which is critical for their transition to deflagration and detonation.[11] Despite significant efforts, many aspects of how these factors intertwine to dictate the sensitivity and performance of an energetic material remain unresolved.

The multi-scale nature of shock-to-deflagration and shock-to-detonation makes them especially challenging to investigate with computational methods, as simulations inherently face a trade-off between physical fidelity and computational scalability. High-fidelity methods like atomistic molecular dynamics (MD) can capture detailed molecular processes but are limited to small length (up to ~micron) and nanosecond time scales. Conversely, lower-fidelity coarse-grain (CG) models can simulate mm scales required for shock to detonation by simplifying the underlying physics. To bridge this gap, machine learning (ML) models have recently emerged as a powerful strategy.[12-16] Examples include interatomic potentials for large-scale MD simulations approaching density functional theory (DFT)[17, 18] accuracy and dimensionality reduction techniques to learn reduced chemical kinetics models.[19] ML tools have also been successfully applied to accelerate physics-based simulations such as the work of Baek and Udaykumar in their physics-aware recurrent convolutional neural networks (PARC) model.[20, 21] PARC incorporates the physics of multi-phase flow in its architecture and is trained with direct numerical simulations. The PARC model has been demonstrated to simulate the shock initiation of polycrystalline energetic materials. Additionally, ML models capable of learning from multimodal data are being employed for multiscale modeling of materials. For example, ML interatomic potentials have been trained with a small amount of high-fidelity data with a larger amount of low-fidelity data (e.g. semilocal DFT), can lead to ML models with multi-scale predictive capability in data-efficient manner.[22, 23] In our example, we combine atomistic simulations and particle-based coarse-grain simulations to simulate the interaction of shockwaves with a wide range of microstructures. These data are used to train a ML model, Microstructure-Informed Shock-induced Temperature net (MISTnet2), capable of mapping microstructure to shock-induced temperature.

MISTnet2 utilizes a conditional 3D U-Net architecture[24] to predict shock-induced temperature fields from all-atom MD simulations of RDX/polystyrene composites and CG dissipative particle dynamics with energy conservation (DPDE) simulations of porous RDX. The model is explicitly conditioned on the simulation type (atomistic or CG), the applied shock particle velocity ($u_p$), and the spatial length scale of the input data. This allows a single network to learn the complex mapping from an initial 3D microstructure to the resultant 3D temperature field across different physical regimes, from $O(10$ nm) scale of the atomistic systems to $O(\mu$m) scale of the CG systems.

## 2. Results

### 2.1 MISTnet2 model architecture

MISTnet2 is a 3D[24] conditional U-Net[25] model designed to predict the shock response of RDX PBXs and porous RDX for varying particle velocities ($u_p$), length scales, and underlying simulation model. The training datasets for the model consisted of atomistic molecular dynamics (MD) simulations of polymer-bonded explosives (PBX, Fig. 1a) and coarse-grain (CG) simulations of crystalline RDX with multiple pores (Fig. 1b). The PBX consisted of 1,3,5-trinitro-1,3,5-triazinane (RDX) nanoparticles (NP) and polystyrene (PS) binders and non-reactive force fields of Smith-Bharadwaj potential[26] (RDX) and Dreiding[27] (PS) were utilized. The PBX microstructures were used in our previous study[28] and denoted to as S1-S7 in Table 1 of that work. Briefly, these slabs had dimensions of 61 nm × 18 nm × 100 nm, density of approximately 1.49 g/cm³, and PS weight percent ranging from 9.5% to 14.7%. The applied piston velocities ($u_p$) were 1.0, 1.5, and 2.0 km/s. We complemented the MD PBX data with CG simulations to reach larger spatial and temporal scales. The CG simulations utilized the energy-conserving dissipative particle dynamics model (DPDE). The microstructures consisted of crystalline RDX with dimensions of 500 nm × 4 nm × 1 μm and pores at varying locations as shown in Fig. 1b. Ten slabs were randomly generated to contained 4 pores with diameter of 100 nm, 10 pores with diameter of 50 nm, and 15 pores with diameters of 20 nm. More detailed descriptions of the simulated datasets are given in the Method Section 4.1.

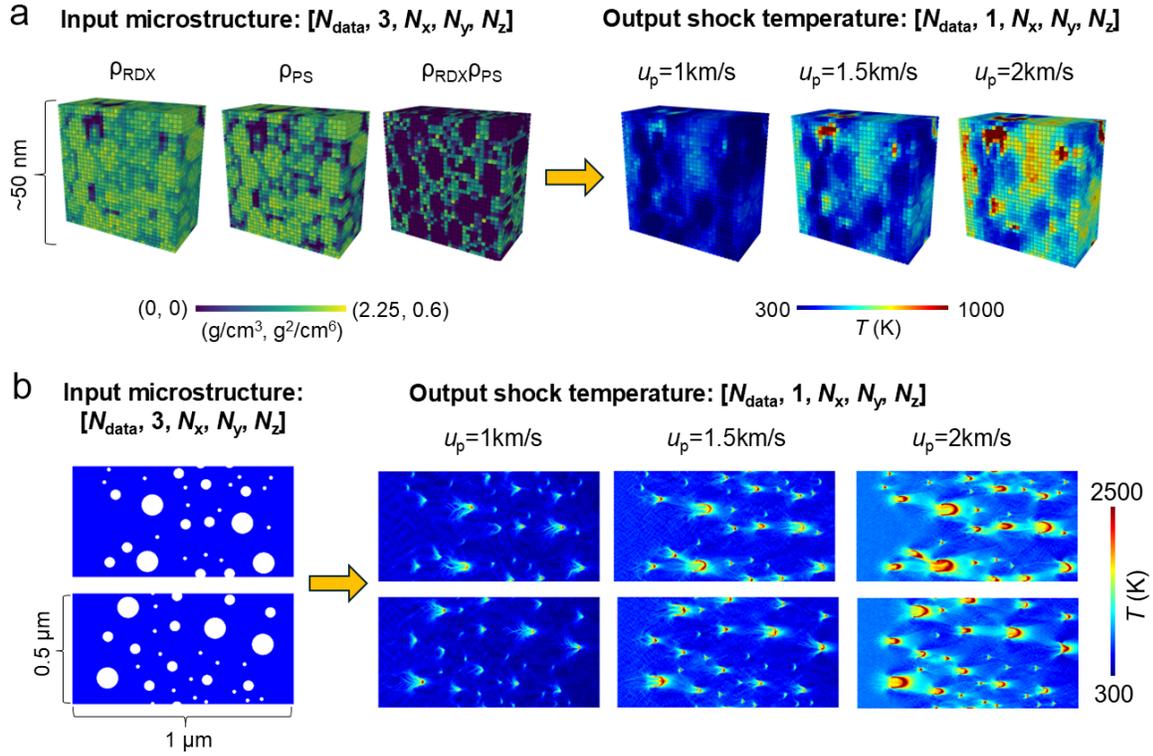

**Figure 1**. Dataset for training MISTnet2. (a) Atomistic MD initial microstructures are converted into spatially binned 3 channeled data that represent the initial densities ($\rho_{RDX}$, $\rho_{PS}$) and their products ($\rho_{RDX}\rho_{PS}$). The outputs are spatially binned temperatures. (b) CG DPDE initial microstructures were pure RDX slabs with randomly generated pores. For both dataset, $u_p$ of 1, 1.5, 2 km/s were applied.

MISTnet2 is designed to map the initial microstructural features of these datasets to the shock response of PBX. The inputs were spatially binned $\rho_{RDX}$, $\rho_{PS}$ and $\rho_{RDX}\rho_{PS}$ with units of g/cm³. In addition, MISTnet2 requires conditional inputs of $u_p$, simulation model used to generate the data, and the spatial resolution of the bins. The spatial resolutions were 3 scalar numbers for the Cartesian coordinates with units of nm and $u_p$ was a scalar number with units of km/s. The conditional input for simulation model was a one-hot vector of size two to indicate whether the data correspond to either atomistic MD or CG DPDE simulation. The output of the model were temperatures in Kelvin divided by 1000. More detailed description of the model is given in the Method Section 4.2.

### *3.2 Model accuracy for atomistic simulations*

As a first test of MISTnet2 accuracy, we evaluated the model performance on the PBX shock response obtained from atomistic MD simulations. The results are depicted in Fig. 2. Hotspots are formed from interaction between the applied shock and various microstructural features, through mechanisms such as pore collapse, plasticity, and interfacial friction between polymer and RDX. Such mechanisms have different dependencies on the shock strength, given by $u_p$, and lead to a wide range of hotspots with temperatures in the range of 500-2500 K. We evaluate the model performance through visual inspection (Fig. 2a, d), cumulative temperature vs. volume (T-V) plots (Fig. 2b, e), and parity plots (Fig. 2c, f). In Section 2.6, we further evaluate the accuracy of MISTnet2 by assessing the agreement between the predicted and ground truth temperature fields concerning whether the hotspots quench or deflagrate. The cumulative T-V displays the volume within the material that is above certain temperature and is used to understand both the size and temperature of the shock-induced hotspots, which is critical to their deflagration.[11] This metric measures the global accuracy of the ML models compared to parity plot that evaluates the pixel-level accuracies.

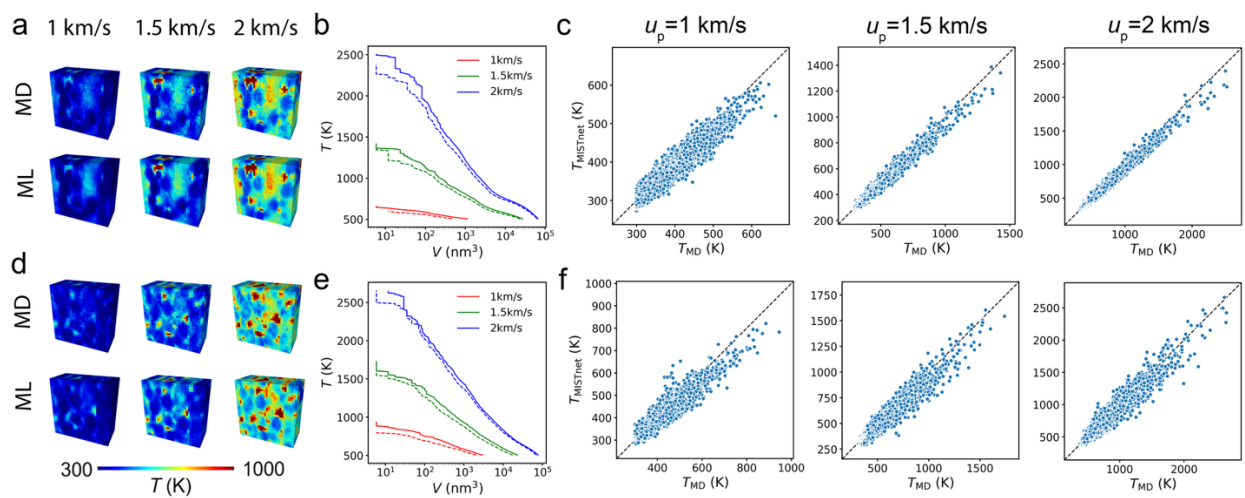

**Figure 2**. Prediction accuracy for atomistic PBX. (a, d) Comparison of temperature fields from atomistic MD simulations and ML predictions. (b,e) Cumulative temperature versus volume. (c,f)

Parity plots of temperatures. (a-c) are for training dataset while (d-f) are for test sets. The solid lines are for ML results and dashed lines are for MD for the cumulative temperature versus volume plots.

Visual inspection of the hotspots show, as was the case in our previous study for unconditional MISTNet,[28] remarkable agreement between MISTnet2 and the MD results. The ML model correctly identifies all hotspots including the small hotspots with the highest temperatures (highlighted by purple rectangles in Fig. 1a) and the large hotspots with moderate temperatures (black rectangles on Fig. 1a). The parity plots also show that the ML model accurately predicts the temperature and location at the individual pixel level. Similarly, the cumulative T-V plots show good agreement between MD and ML for both training and test datasets for all $u_\mathrm{p}$. These results show that the conditional ML architecture is sufficient to achieve highly accurate ML model that was previously established for one $u_\mathrm{p}$ with unconditional U-Net.

### 2.3 Model accuracy for CG simulations

The model performance for varying length scales and simulation methods was evaluated using CG multi-pore simulations. The corresponding microstructures differ significantly from the atomistic PBX, as they consist of pure RDX crystals with pore diameters ranging from 20 to 100 nm. We note that the original MISTnet was only able to qualitatively predict the hotspot associated with a single pore embedded in RDX and the quantitively erred by $O(1000\ \mathrm{K})$ for the maximum temperatures (see Supplementary Materials Figure S11 from Li et al.[28]).

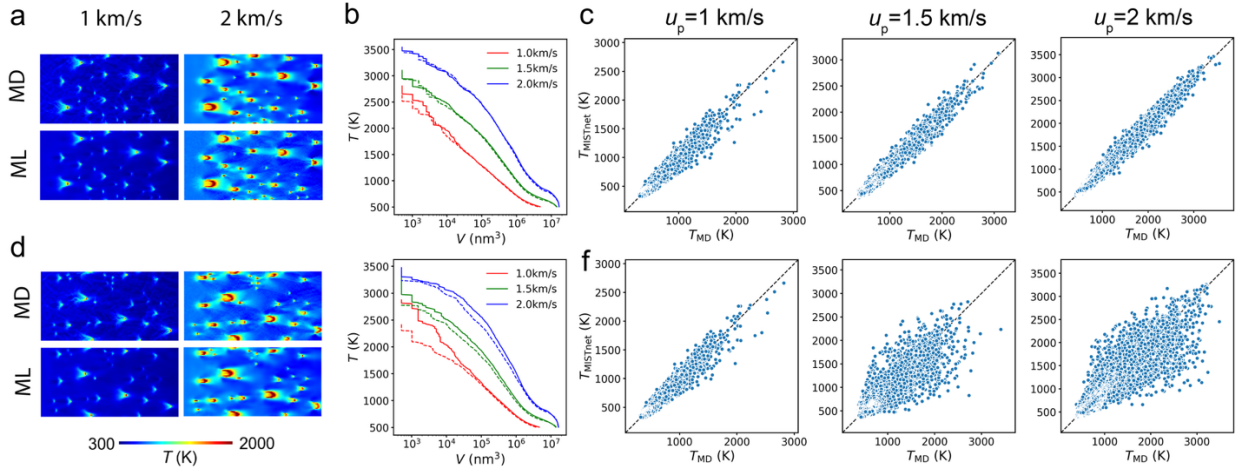

**Figure 3**. Prediction accuracy for CG porous RDX. (a, d) Comparison of temperature fields from CG DPDE simulations and ML predictions. (b,e) Cumulative temperature versus volume. (c,f) Parity plots of temperatures. (a-c) are for training dataset while (d-f) are for test sets.

Similar to our analysis of the atomistic PBX results, we evaluate the model performance through visualization (Fig. 3a, Fig. 3d, Fig. 4), cumulative T-V plots (Fig. 3b, Fig. 3e) and parity plots (Fig. 3c, Fig. 3f). For both train and test datasets, the cumulative T-V plots show good agreement between the ML and DPDE results, demonstrating overall high accuracy of ML predictions. However, the parity plots show reduced correlation between the DPDE and ML results for the test set in comparison to the training dataset where the parity plots are displaying nearly one-to-one correspondence between the two methods.

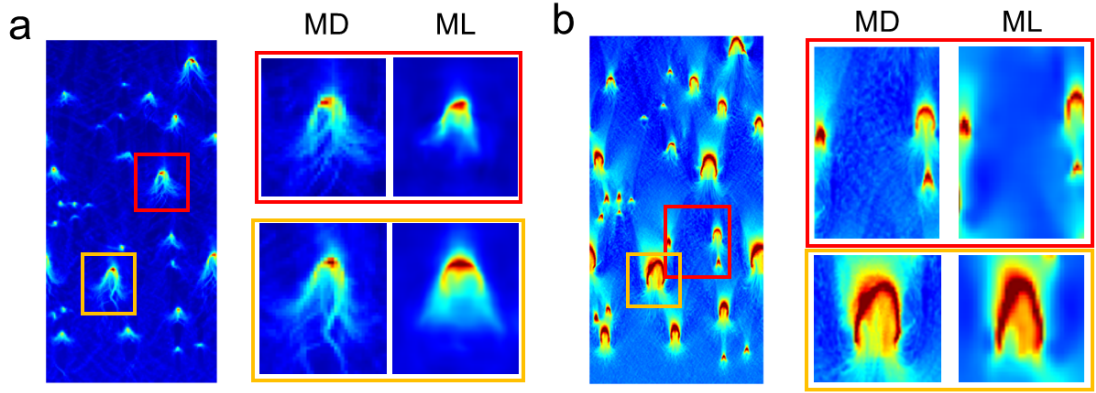

**Figure 4**. Detailed comparison of hotspot morphologies from DPDE (left panels) and ML predictions (right panels) for the test dataset at (a) $u_p$=1 km/s and (b) $u_p$=2 km/s.

A magnified view of the hotspots depicted in Fig. 4 explains the reduced pixel-level accuracy. As the CG materials are multi-pore slabs with some of the pores in close proximity to each other, there are multi-pore interactions that affect the morphology of the hotspots. For example, many of the downstream pores are not impacted by shock waves moving parallel to the piston, as the shock focusing occurs on upstream voids and shear bands are formed from stress concentration. This leads to skews and diverse shapes of hotspot tails. Such effects are captured by MISTnet2 as demonstrated by the hotspot morphology depicted in Fig. 4b, region enclosed by an orange box. However, such complexity is more difficult to predict with pixel-level accuracy and leads to the less correlated parity plots for the test set as depicted in Fig. 3f. This effect becomes more pronounced as the bin resolutions become finer as demonstrated by the parity plots in SI Fig. S1. However, the size and temperature level of the hotspots, which are more critical for predicting the deflagration conditions, are accurately captured by the ML model as demonstrated by the cumulative T-V plots on Fig. 3e.

*2.4 Model accuracy for interpolation and extrapolation of $u_p$*

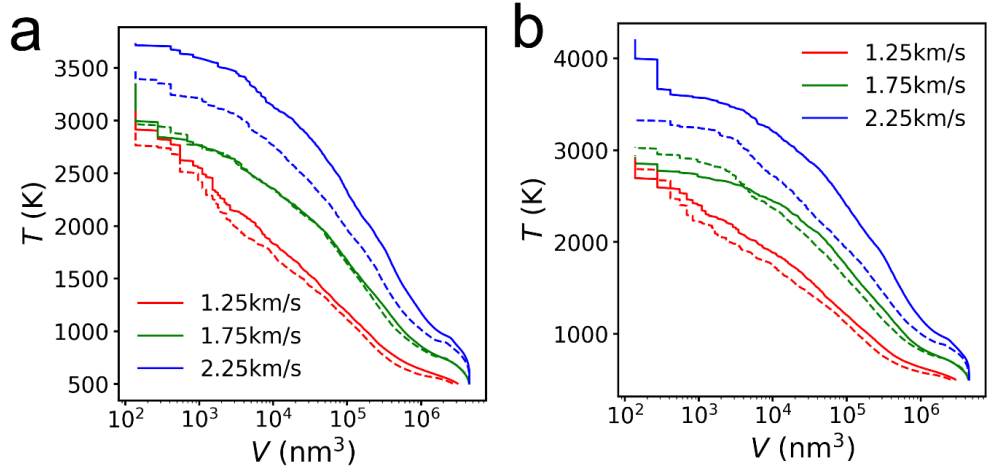

**Figure 5**. Model performance on interpolation and extrapolation of $u_p$ for CG simulations. (a) Microstructure in the training dataset and (b) microstructure in the test dataset.

As another test of MISTnet2's capability, we evaluated its performance for shock strengths ($u_p$) not included in the training dataset for the CG DPDE. Specifically, $u_p$ of 1.25 km/s and 1.75 km/s were examined to evaluate interpolation and $u_p$ of 2.25 km/s was examined to evaluate the model's ability to extrapolate. We evaluated the model on microstructures that were included in the training dataset (Fig. 5a) and in the test dataset (Fig. 5b). However, we stress that all results in Fig. 5 are test set data that were not in the training data. The results show that the model is highly accurate for the interpolation regime (1.25 km/s and 1.75 km/s) for both microstructures. For extrapolation to higher $u_p$, the model correctly predicts higher temperatures but underestimates the effect. These results suggest that the model qualitatively agrees with the physics of the system, but is not quantitatively accurate for conditions that are outside the distribution of the training dataset.

*2.5 Multimodality of the model*

The ability of MISTnet2 to assess the differences in temperature fields predicted by the two physics-based models (MD and CG) was evaluated by observing predicted temperatures for equivalent microstructures.

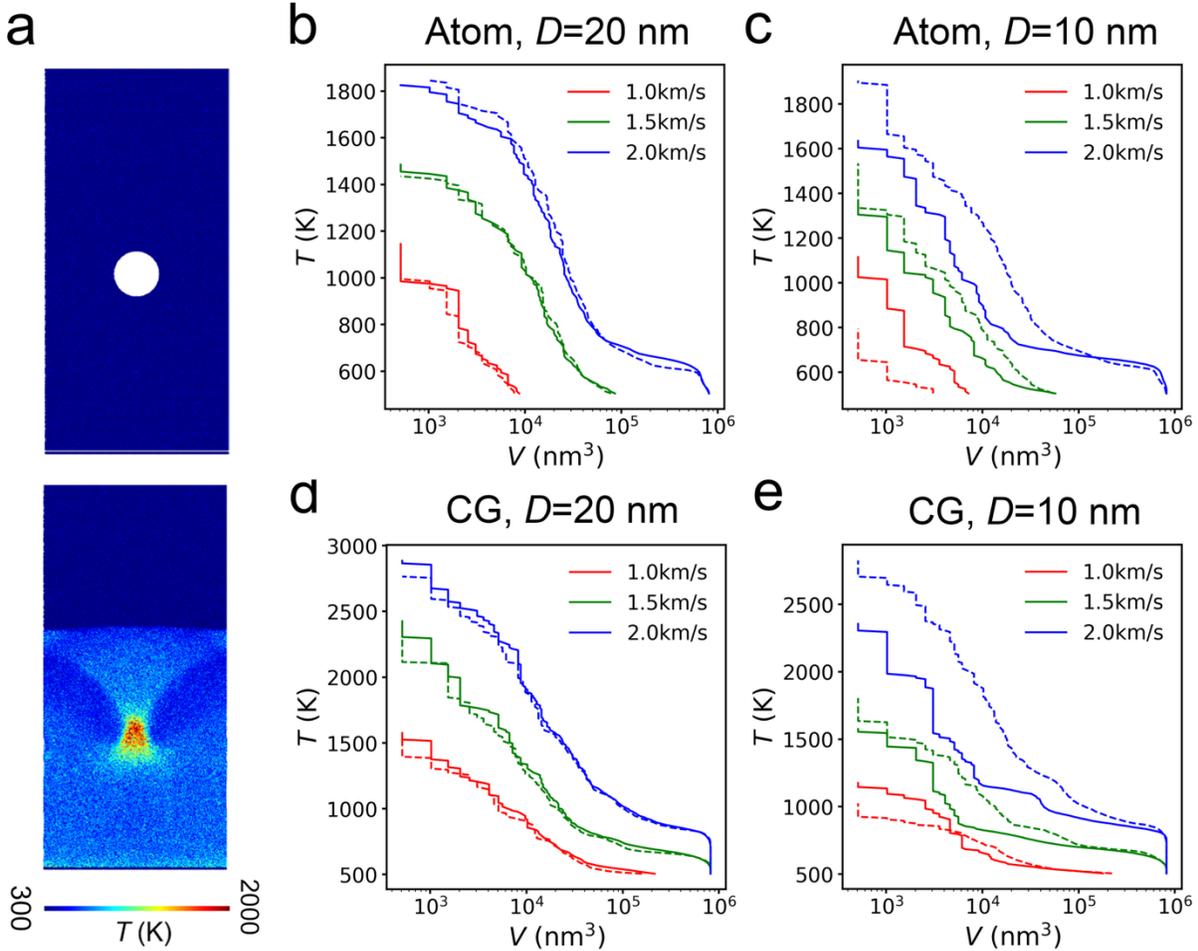

**Figure 6**. Model performance on single pore simulations. (a) Snapshot of single-pore initial structure and shock response. (b-e) Cumulative T-V plots for (b) atomistic, *D*=20 nm, (c) atomistic, *D*=10 nm, (d) CG, *D*=20 nm, and (e) CG, *D*=10 nm. Solid lines are for MD and CG simulations and dashed lines represent MISTnet2 results.

We performed CG and atomistic MD simulations for pure RDX slabs with single large pores with diameters (*D*) ranging from 10 to 20 nm as depicted in Fig. 6a. The cumulative T-V results on Fig. 6 demonstrate that the temperatures differ significantly depending on the simulation model even for the same microstructures, with the CG model leading to maximum temperatures that are approximately 50% higher than the atomistic MD. Several factors contribute to this difference. For example, the CG DPDE utilizes a quantum mechanically derived specific heat to describe RDX molecules, while atomistic MD leads to classical heat capacity. The lack of quantum ionic effects in the MD results in a significant overestimation of the specific heat at low temperatures, which can result in a significant underestimation of the hotspot temperatures, as discussed by Hamilton et al.[29] In addition, the CG RDX potential utilized in this study does not properly describe the shear-induced plasticity and leads to structural differences with the all-atom MD[30-32] that can lead to quantitative difference for hotspot temperatures formed by pore collapses.

For the training datasets depicted on Fig. 6b and Fig. 6d, the ML model is highly accurate. For the test sets on Figs. 6c and. 6e, we see that the ML predictions are accurate for $u_p$=1.5 km/s but over- and underestimate the temperatures at 1km/s and 2km/s. However, the ML predicted temperatures are within reasonable estimation of the simulations and the ML model correctly identifies that CG temperatures should be higher than atomistic results. These results demonstrate that the model qualitatively and somewhat quantitatively captures the general effect of underlying simulation method in the resulting temperatures.

### 2.6 Deflagration prediction from ground truth vs MISTnet2 temperature fields

Among the major objectives of MISTnet2 is to help assess whether a given microstructure will result in deflagration when shocked at a given strength. Thus, to further test the accuracy of MISTnet2, we evaluate whether the predicted temperature field deflagrates or quenches and compare the results with those of the ground truth fields. To predict the post-shock evolution of the temperature fields we used a reduced order chemical kinetics model derived from reactive MD simulations for RDX[19] and thermal transport. The thermal kinetics finite differences algorithm evolves a three-dimensional temperature volume, represented as a simple grid, using the coupled heat diffusion and Arrhenius kinetics equation to solve for the temporal and spatial changes in temperature across time, capturing both the thermal and chemical processes influencing deflagration. The governing equation is:

$$\rho C_v \dot{T} = k \nabla^2 T - Q_1 \dot{C}_1 + Q_2 \dot{C}_3$$

where $\rho$ represents the derived material density of 2.15 g/cm$^3$,[33] classical specific heat is represented as $C_v$ using the value 0.563 $\frac{\text{cal}}{\text{g} - \text{K}}$, thermal conductivity is represented by $k$ using the value 0.361$\frac{W}{\text{m} - \text{K}}$, $Q_1$ and $Q_2$ represent the heat of reaction coefficients, and $\dot{C}_1$ and $\dot{C}_3$ represent the rate of reaction for the reactant and product contributions respectively. The thermal evolution is solved using a timestep of 0.1ps for up to 30 ns or until the average system temperature plateaus (the temperature range is less than 1000 K, and the average temperature has not changed more than 0.05 K) for over 0.5ns. To align with the boundary conditions in the MD simulations, periodic boundary conditions are applied across the shock perpendicular directions, and zero-Neumann boundary conditions are applied in the shock parallel direction. The temperature volumes evolve following a correlated propagating shock wave that acts as a zero-Neumann boundary, where temperature evolution is activated in the region behind the shockwave and frozen ahead of the shockwave. Further details of the implementation are available in the SI section 2. We illustrate the simulations with snapshots of the process for a case that leads to a quenching (top row of Fig. 7) and deflagration (bottom row of Fig. 7). The initial temperature fields correspond to MISTnet2 predictions for the same microstructure from the test set and $u_p = 1.5$ km/s (top row) and $u_p = 2.0$ km/s (bottom). While the 1.5 km/s field quenches, the strong shock results in a predicted deflagration, the final snapshot shown in Fig. 7 highlights the growing deflagration wave that consumes the entire simulation cell at later times.

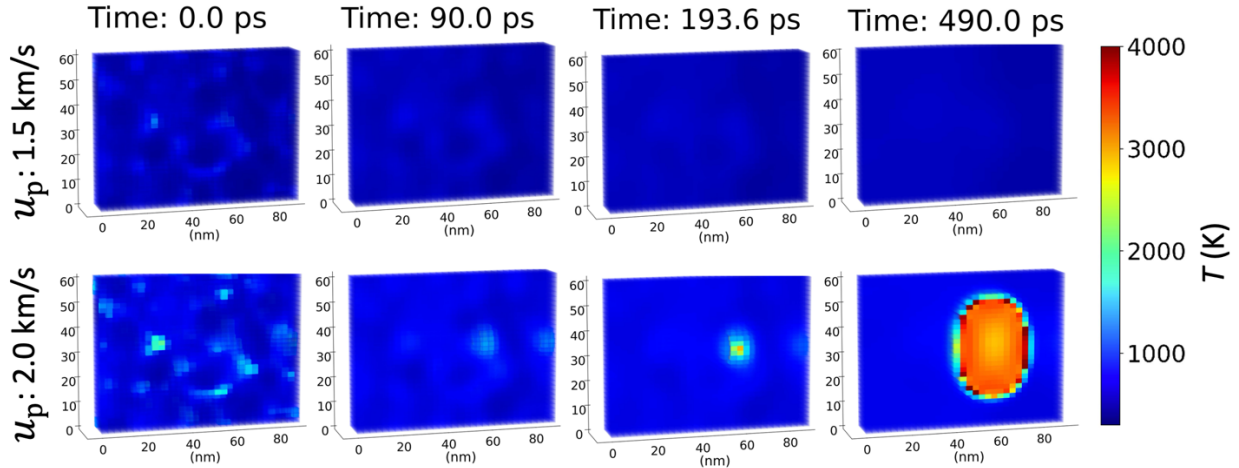

**Figure 7**. Temperature evolution of initial MISTnet2 predicted fields for the same microstructure from the test set and $u_p$ = 1.5 km/s (top) and $u_p$ = 2.0 km/s (bottom).

To assess the accuracy of the ML model, we compare the final temperature obtained by propagating the temperature fields of MISTnet2 and MD for all the PBX and multi-pore microstructures in the testing set for $u_p$ = 1.0, 1.5, and 2.0 km/s. Out of the 30 fields, 15 resulted in a deflagration and 15 quenched. Of the corresponding MISTnet2 fields, 17 deflagrated while the remaining 13 quenched. The disagreement lies in the PBX $u_p$ = 2.0 km/s test set, where 5 of the 6 MISTnet2 fields deflagrate while 3 of the 6 MD fields deflagrate. Figure 8 further breaks down the results. 100% of the MD and MISTnet2 multi-pore temperature fields are reactive, concluding complete deflagration agreement for this dataset. Across the PBX systems, 100% of the MD and MISTnet2 temperature fields quench at the 1.0 and 1.5 km/s particle velocities; however, 21.4% (9/42) of the MD PBX temperature fields deflagrate at the 2.0 km/s particle velocity whereas 33.3% (14/42) of the MISTnet2 PBX temperature fields deflagrate. This leads to an overall deflagration agreement of 98.4% (301/306) and 88.1% (37/42) for the PBX systems shocked at 2.0 km/s particle velocity, where 87.5% (21/24) of the training set agree, 100% (12/12) of the validation set agree, and 66.7% (4/6) of the test set agree.

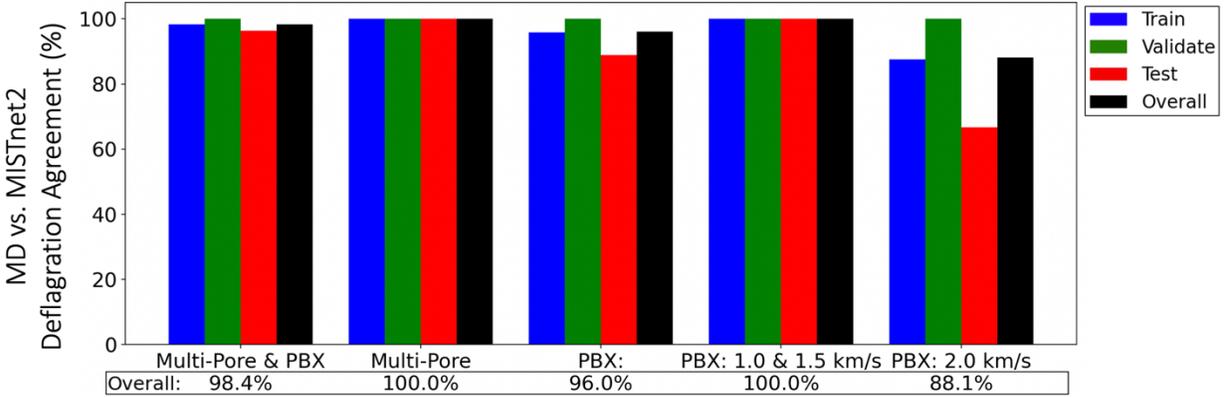

**Figure 8**. Deflagration agreement between the MD and MISTnet2 multi-pore and PBX systems.

## 3. Discussion

We introduced MISTnet2, a model based on the 3D U-Net architecture capable of predicting shock-induced temperature fields in a wide range of microstructures of RDX energetic formulations and for a wide range of shock strengths. By conditioning the network on key physical parameters, such as applied piston velocity, underlying simulation model, and spatial resolution, we have demonstrated a significant advancement in versatility over the previous state of the art (unconditional MISTnet). Our results show that MISTnet2 model can accurately predict the complex 3D temperature fields resulting from the interaction of shockwaves with wide range of microstructures, physical regimes and material representations.

The model demonstrated excellent quantitative accuracy for systems within its training data distribution, including complex cases involving multi-pore interactions, polymers-RDX interfaces, and interpolated simulation conditions. However, for extrapolation into conditions outside the training dataset, the model generally only captures the qualitative trends in hotspot temperatures and the numerical accuracy is not maintained. Future efforts can focus on expanding the training dataset to improve extrapolation and include a wider variety of material properties, such as the shock-induced pressures or strains. Overall, by drastically reducing the computational cost of inference from thousands of CPU hours to mere GPU seconds, MISTnet2 provide a method for the rapid design and analysis of energetic materials with tailored safety and performance characteristics.

# 4. Methods

## 4.1 Data generation

MISTnet2 was trained using shock simulations using atomistic and CG simulations of shocked PBX systems and single crystals with porosity. The methods for the atomistic molecular dynamics (MD) simulations were outlined in our previous study[28] and will be discussed briefly here. Both MD and CG simulations were performed using the open-source software LAMMPS.[34]

The atomistic PBX microstructures consisting of RDX and PS were prepared using the PBXgen code described in Li et al.[35], an in-house tool for preparing PBX initial structures. The prepared PBX systems were replicated along the shock direction ($z$) and dynamic loading was applied from the top and bottom surfaces, leading to two independent shock temperature fields. To train MISTnet2, we augmented these shock temperature results by applying rigid rotations to the input and output data in the axes normal to the shock directions. Therefore, each simulation led to six data sets (top, bottom, top flipped in the $x$ direction, top flipped in the $y$ direction, bottom flipped in the $x$ direction, bottom flipped in the $y$ direction). The piston velocities included in training are 1 km/s, 1.5 km/s, and 2km/s. In total, we generated and used 126 new atomistic MD data sets compared to the original MISTnet work. To create input for the MISTnet2, we binned the initial microstructures into $32 \times 16 \times 32$ spatial bins of sizes approximately 1.9 nm $\times$ 1.1 nm $\times$ 2.8 nm. We describe the initial microstructure using three descriptors. For each bin, we measure the density of RDX ($\rho_{RDX}$), density of PS ($\rho_{PS}$), and their product ($\rho_{RDX}\rho_{PS}$) that signals the interfacial area of RDX and PS. The output was the shock temperature after shock has passed through the material. A representative atomistic dataset is given in Fig. 1a and snapshots all the training and testing data are include in the Supplementary Information (SI), Section 4.

The CG model simulations used energy-conserving dissipative particle dynamics (DPDE). We refer to Lísal et al.[36] for a detailed description of the method. The RDX molecules in this framework are described as a single CG bead whose intermolecular conservative forces are described as isotropic forces developed by Moore et al.[37] This DPDE RDX model has been used to study its shock physics[38-41] and have demonstrated reasonable agreement with atomistic models in describing the shock induced hotspot formation.[42] The RDX slabs have dimensions of 500 nm × 4 nm × 1 μm with pores at varying locations as shown in Fig. 1b. Each slab contained 4 pores with diameter of 100 nm, 10 pores with diameter of 50 nm, and 15 pores with diameters of 20 nm. Ten randomly generated CG microstructures were used for the training and testing of MISTnet2. The pore locations were chosen from uniform random number generation that excluded overlapping positions between the pores. The applied $u_p$'s were 1 km/s, 1.5 km/s, and 2 km/s. We utilized three resolutions (2 nm, 4 nm, and 8 nm) of the spatial bins in the $x$ and $z$ dimensions to create the input and output for the MISTNet2, leading to 90 data points. The results in the main manuscripts are from resolutions of 4 nm and the results for 2 nm and 8 nm are in the SI section 1. The CG slabs were quasi-two-dimensional in the $y$ dimension, so we binned them as single bins in $y$ dimensions and replicated them eight times for numerical stability. Furthermore, as the CG microstructures were from pure RDX crystals without PS, the $\rho_{PS}$ and $\rho_{RDX}\rho_{PS}$ input to MISTnet2 were zero.

## 4.2 Conditional U-Net and training scheme

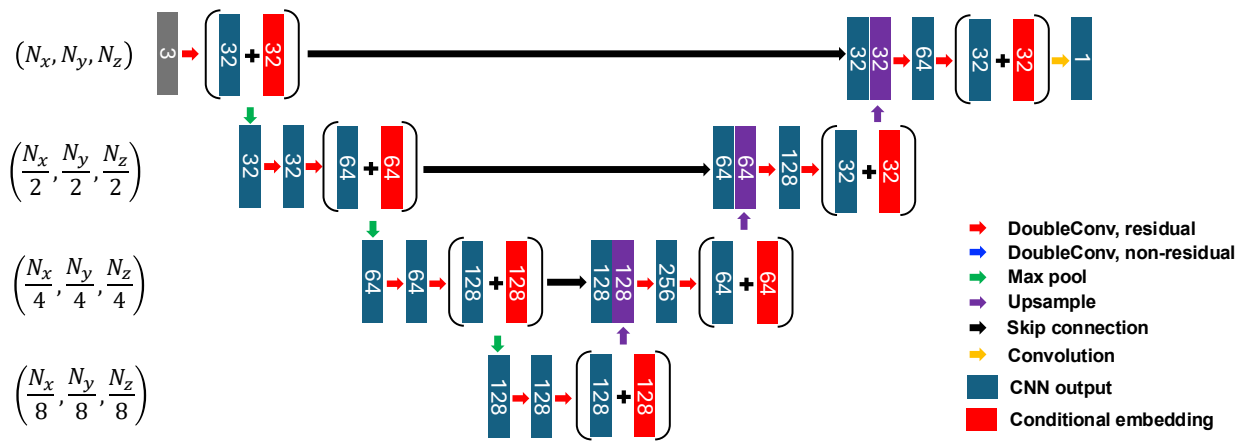

**Figure 9**. Architecture of MISTnet2. $N_x$, $N_y$, and $N_z$ represent the initial input dimensions and the values within the squares represent the number of channels.

The conditional U-Net consisted of encoder, decoder, and skip connections as depicted in Fig. 9. The encoder comprises four *down* blocks that expand the channel dimensions 3→32→64→128→128 through two *DoubleConv* modules while halving the spatial resolutions through max pooling operations. Each *DoubleConv* module uses two successive $3 \times 3 \times 3$ convolutions with bias disabled, kernel size of three and a stride of one. Each convolution is followed by group normalization[43] and GELU[44] activation. This module has a residual option that adds the block's input to its processed output before the final activation as,

$$\mathbf{x}_{\text{out}} = \text{GELU}\big(\mathbf{x}_{\text{in}} + \text{DoubleConv}(\mathbf{x}_{\text{in}})\big) \tag{1}$$

In each *down* blocks, the first *DoubleConv* is applied with a residual option and the second module is without the residual option.

The decoder mirrors the encoder and consists of three *up* blocks that first concatenate the up-sampled feature map with its encoder skip connections. This converts the channels to 256→128→64→32. The outputs of all *down* and *up* blocks are modified by conditional inputs following feature-wise linear modulation strategy.[45] The conditional inputs go through two-layer neural networks that convert them to vectors of size 256 followed by ReLU activations to produce embeddings. For every encoder and decoder stage, these embeddings are transformed by another embedding layers that consist of SiLU activation and linear layer that maps 256 to *C*, where *C* is the block's channel width. These embeddings are broadcast across all spatial dimensions and added to the outputs of down and up blocks.

The models were trained for 500 epochs with the AdamW optimizer.[46] The dataset was split approximately 60/20/20 for the training/validation/test sets, respectively. The learning rate was $3\text{x}10^{-4}$. Similar to our previous paper,[28] we use weighted mean-squared error (WMSE) for the loss function. For the simulation ($y$) and MISTnet2 predicted ($\hat{y}$) temperatures in Kelvin normalized by constant division of 1000, we use following loss function,

$$\text{WMSE} = \frac{1}{\text{N}} \sum_{i=1}^{N} w(y_i)(y_i - \hat{y}_i)^2 \,; w(y) = \begin{cases} 10, y \geq 6 \\ 5, 3 \leq y < 6 \\ 1, y < 3 \end{cases} \tag{2}$$

Here, $N$ is the number of spatial bins in each data point. We use these weighted loss functions because the current dataset contains significantly higher temperatures compared to our previous MISTnet dataset as the CG microstructures contained larger pores.

## Code availability

MISTnet2 is implemented using PyTorch and the SI section 3 includes pseudocode with all the required details for its reproduction.

## Supplementary information

The supplementary information includes bin size dependent model performance, detailed explanation of the thermos-chemical simulations, pseudocode for the MISTnet2, and images of all training and testing sets to enable to complete evaluation of the model.

## Acknowledgments

This research was sponsored by the Army Research Laboratory and was accomplished under Cooperative Agreement Number W911NF-20-2-0189. This work was supported in part by high-performance computer time and resources from the DoD High Performance Computing Modernization Program. The views and conclusions contained in this document are those of the authors and should not be interpreted as representing the official policies, either expressed or implied, of the Army Research Laboratory, or the U.S. Government. The U.S. Government is authorized to reproduce and distribute reprints for Government purposes notwithstanding any copyright notation herein.

## Data availability

The supplementary information includes snapshots of all training and testing sets to enable to complete evaluation of the model. The data that support the findings of this study are available from the corresponding author upon reasonable request.

## Author declarations

The authors have no conflicts to disclose.

# Supplementary Information: Multi-Fidelity Predictive Model for Shock Response of Energetic Materials Using Conditional U-Net

Brian H. Lee,[1] Chunyu Li,[1] Aidan Pantoya,[1] James P. Larentzos,[2] John K. Brennan,[2] Alejandro Strachan[1, a)]

Author affiliations
[1]*School of Materials Engineering and Birck Nanotechnology Center, Purdue University, West Lafayette, Indiana 47907, USA*
[2]*U.S. Army Combat Capabilities Development Command (DEVCOM) Army Research Laboratory, Aberdeen Proving Ground, Maryland 21005, USA*

Author email
[a)]Author to whom correspondence should be addressed: strachan@purdue.edu

This document contains the following information.

**1. Bin resolution dependent coarse-grained model results**. This compares the accuracy of MISTnet2 for inputs with different bin sizes.

**2. Thermo-chemical simulations to evaluate the accuracy of MISTnet2.** This section includes additional results.

**3. MISTnet2 pseudocode**. MISTnet2 is implemented using PyTorch the pseudocode describes the main data workflow, model architecture, and objective function.

**4. Training and testing images**. Images associated with all training, validation, and testing datasets. They include input channels and output channels. These data include, total density, RDX density, and the product of the RDX and polymer densities (inputs) and final temperature following shock loading. We stress that, while capturing realistic features, the microstructures are synthetic and do not represent any real formulation.

## Data Release Statement:

MISTnet2 is trained on artificially generated datasets that, while representing physical characteristics, do not correspond to any real or sensitive material. The data effectively evaluates the proposed methods. The MISTnet2 model is an implementation of commonly accessible convolutional network architectures and poses no sensitivity risk to open release.

# 1. Bin resolution dependent coarse-grained model results

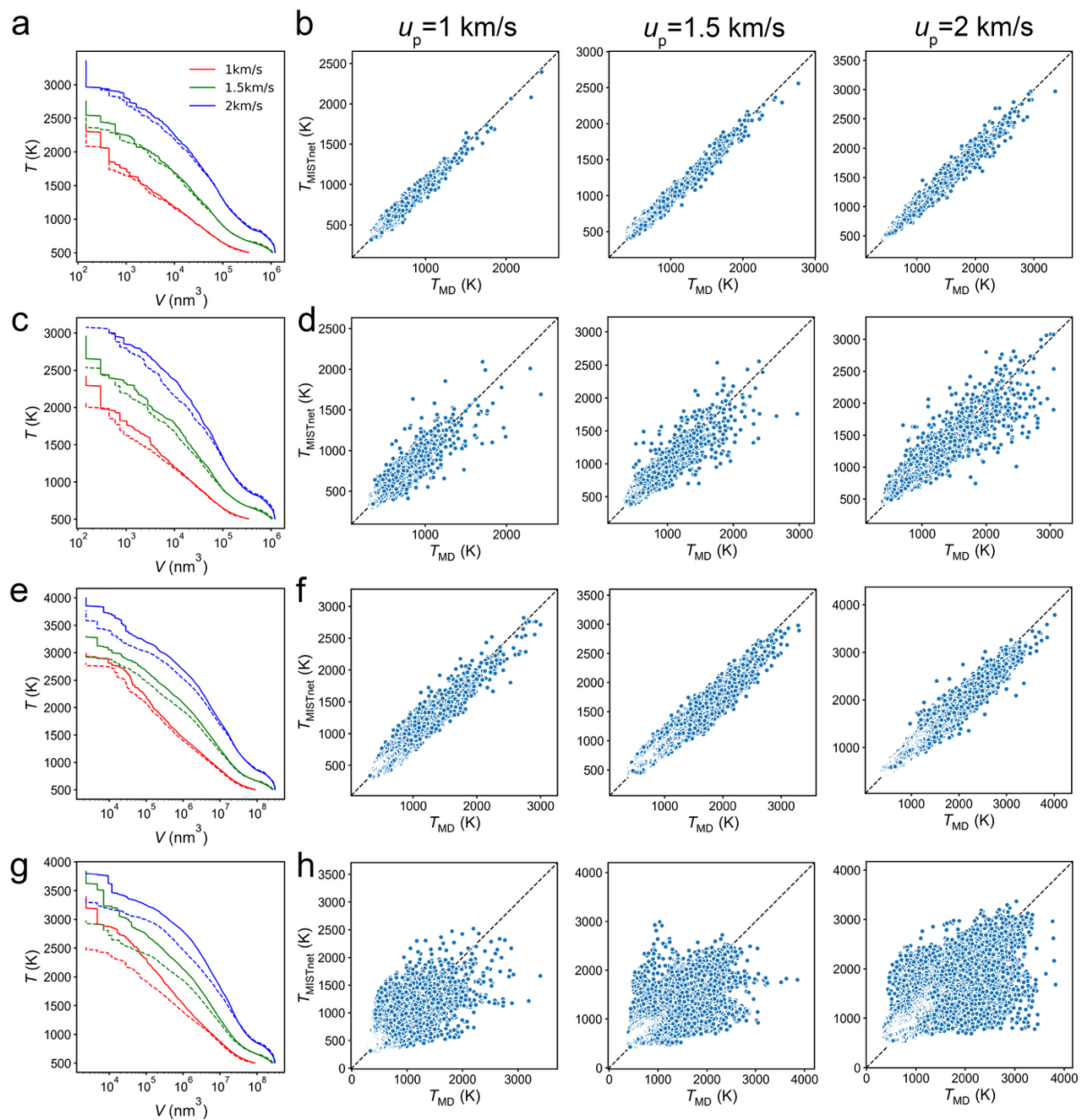

**Figure S1**. Prediction accuracy for CG porous RDX. (a, c, e, g) Cumulative temperature versus volume. (b,d,f,h) Parity plots of temperatures. (a, b) are for training dataset and (c, d) are for test dataset with bin resolution of 2 nm. (e, f) are for training dataset and (g,h) are for test dataset with bin resolution of 8 nm.

# 2. Thermo-chemical continuum model and additional results

The thermal continuum model used in predicting the deflagration of hotspots is an implementation of the coupled heat diffusion and Arrhenius kinetics equations, parameterized from reactive molecular dynamics (MD). The governing equation is represented as:

$$\rho C_v \dot{T} = k\nabla^2 T - Q_1 \dot{C}_1 + Q_2 \dot{C}_3$$

This multi-process equation is evaluated in three major steps: first calculate heat diffusion, second calculate Arrhenius reaction kinetics, and third update the temperature field with heat diffusion and Arrhenius reaction kinetics simultaneously. Heat diffusion is represented as:

$$\rho C_v \dot{T} = k\nabla^2 T$$

Density ($\rho$) is 2.15 (g/cm$^3$). Specific heat capacity ($C_v$) is 0.563 ($\frac{\text{cal}}{\text{g} \cdot \text{K}}$). The time derivative of temperature, rate of change of temperature with respect to time, ($\dot{T}$) is solved with:

$$\dot{T} = \frac{1}{\rho C_v}(k\nabla^2 T - Q_1 \dot{C}_1 + Q_2 \dot{C}_3)$$

Thermal conductivity ($k$) is 0.361 ($\frac{\text{W}}{m \cdot \text{K}}$). The Laplacian of temperature, spatial temperature second derivative, ($\nabla^2 T$) is calculated using the second-order central finite difference approximation (for every direction: x, y, z):

$$\frac{d^2 T}{d_x{}^2} \approx \frac{T_{i+1} - 2T_i + T_{i-1}}{(\Delta x)^2}$$

X-Dimension implementation in python script:

d2T_dx2[1:-1, :, :] = (T[2:, :, :] - 2 * T[1:-1, :, :] + T[:-2, :, :]) / dx**2 # Internal Points

d2T_dx2[0, :, :] = (T[1, :, :] - 2 * T[0, :, :] + T[1, :, :]) / dx**2 # Boundary Zero Neumann

Boundaries parallel to the shock are closed and fixed with zero-heat flux (Zero Neumann Boundary Conditions). The shockwave itself is represented with a zero-heat flux boundary, where the region that the shockwave has passed in activated and the region ahead of the shockwave is frozen. Boundaries perpendicular to the shock are periodic. The Arrhenius reaction kinetics portion of the partial differential equation is described:

$$\rho C_v \dot{T} = -Q_1 \dot{C}_1 + Q_2 \dot{C}_3$$

Heat of reaction coefficients parameterized from MD simulations ($Q_1$) and ($Q_2$):

$$Q_i(T) = \begin{cases} a_1 & \text{if } T \leq T_{\text{ref}} \\ a_1 + b_1(T - T_{\text{ref}}) & \text{if } T \geq T_{\text{ref}} \end{cases}$$

These coefficients are calculated with:

$$Q_1(T) = \begin{cases} -1.7 & \text{if } T \leq 1736.3\text{K} \\ -1.7 + 0.0314(T - 1736.3) & \text{if } T \geq 1736.3\text{K} \end{cases}$$

$$Q_2(T) = \begin{cases} 416 & \text{if } T \leq 1736.3\text{K} \\ 416 - 0.03836(T - 1736.3) & \text{if } T \geq 1736.3\text{K} \end{cases}$$

The rates of chemical reaction ($\dot{C}_i$) are calculated as:

$$\dot{C}_1 = -C_1 Z_a \exp(-\frac{E_a}{RT})$$

$$\dot{C}_2 = C_1 Z_a \exp(-\frac{E_a}{RT}) - C_2 Z_b \exp(-\frac{E_b}{RT})$$

$$\dot{C}_3 = C_2 Z_b \exp(-\frac{E_b}{RT})$$

The first order reactions, representing molar fractions of reactants ($C_1$), intermediates ($C_2$), and products ($C_3$) are initialized at each system coordinate with $C_1 = 1, C_2 = 0, C_3 = 0$ (representing the molar fractions).

Then, for i=0 to i=n, by Δtime-step (s) {

$$\dot{C}_1 = -C_1 Z_a \exp(-\frac{E_a}{RT})$$

$$\dot{C}_2 = C_1 Z_a \exp(-\frac{E_a}{RT}) - C_2 Z_b \exp(-\frac{E_b}{RT})$$

$$\dot{C}_3 = C_2 Z_b \exp(-\frac{E_b}{RT})$$

$$C_1 = C_1 + \dot{C}_1 \cdot \text{Δtime-step (s)}$$

$$C_2 = C_2 + \dot{C}_2 \cdot \text{Δtime-step (s)}$$

$$C_3 = C_3 + \dot{C}_3 \cdot \text{Δtime-step (s)}$$

Update Temperature field here with Arrhenius kinetics and heat diffusion simultaneously.}

The remaining Arrhenius kinetics variables have been parameterized and are defined:

$\ln(Z_a) = 31.1 \pm 0.4$: modeled using 31.1 ($\text{s}^{-1}$)

$\ln(Z_b) = 30.1 \pm 0.4$: modeled using 30.1 ($\text{s}^{-1}$)

$E_a = 24.4 \pm 1.5$: modeled using 24.4 (kcal/mol)

$E_b = 23.5 \pm 1.6$: modeled using 23.5 (kcal/mol)

$R = 8.314$ (kcal/mol·K): Universal Gas Constant

After calculating the diffusion and reactivity terms, temperature can be updated:

$$\dot{T} = \frac{1}{\rho C_v}(k\nabla^2 T - Q_1\dot{C}_1 + Q_2\dot{C}_3)$$

$$T = T + \dot{T}$$

This continuum partial differential equation is implemented through Python.

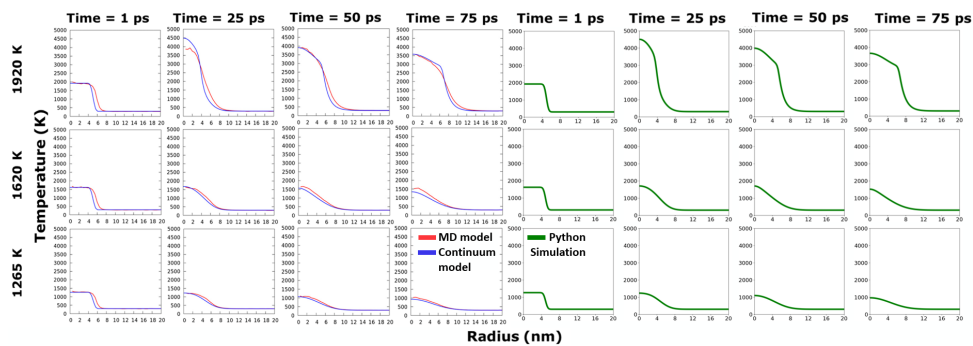

**Figure 2.** The thermal kinetics model used to predict deflagration (green) compared to Sakano's original implementation (blue) and Sakano's reactive MD modeling (red).

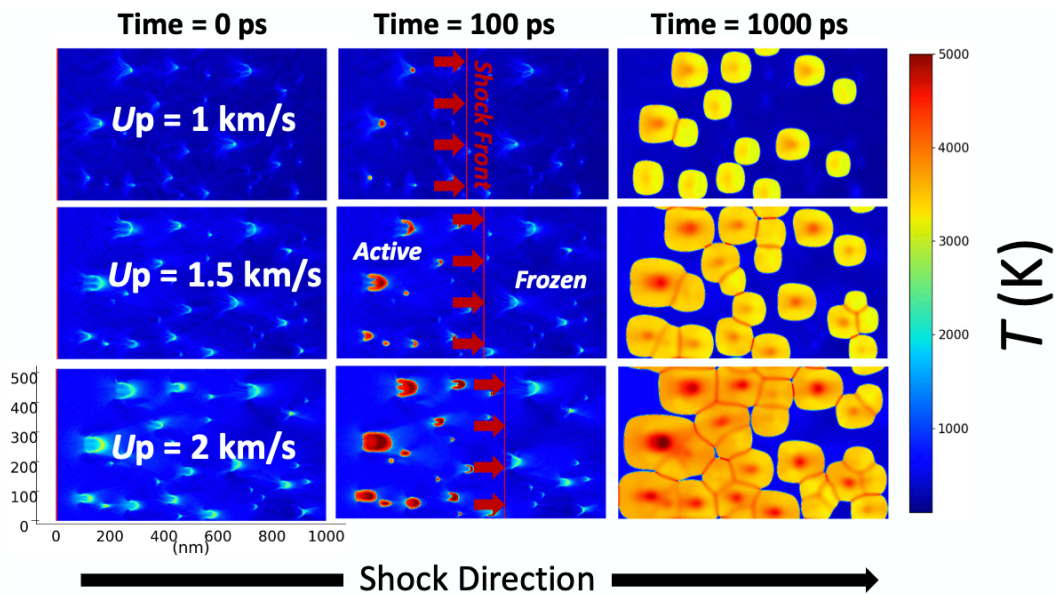

**Figure 3.** Temperature evolution of a multi-circular-pore system shocked at 1 km/s (top), 1.5 km/s (middle) and 2 km/s (bottom). The temperature field evolved following the shock wave, consistent to what would physically be observed.

# 3. MISTnet2 Model Pseudo Code:

MISTnet2 is a 3D U-Net that utilizes the microstructure of a polymer bonded explosive (PBX) as input and the voxel dimensionality, simulation type (atomistic or dissipative particle dynamics), and shock strength as the conditional input. The output of the model is the post-shock temperature field. This is implemented in Python using the PyTorch module. Below we share the pseudocode for our model, which is a standard conditional 3D U-Net, and explanation of the training and evaluation procedure.

**Training and Evaluation Sequence:**

1. Load the training datasets containing the input microstructures (densities and their products given by: $\rho_{RDX}, \rho_{ps}, \rho_{RDX} * \rho_{ps}$), voxel dimensionalities (length of the voxels: $L_x, L_y, L_z$), simulation types, shock strengths (particle velocity, $u_p$), and ground-truth temperature fields into PyTorch DataLoaders. The microstructures and temperatures are 3D tensors. The simulation type is a one-hot vector indicating whether the simulation system is molecular dynamics (MD) or dissipative particle dynamics with energy conservation (DPDE). The other conditional inputs are scalars.

2. Load the validation datasets with the same information into additional PyTorch DataLoaders.

3. Initialize the MISTnet2 model.

4. Create an AdamW optimizer with a learning rate of 3e-4.

5. Set MISTnet2 to training mode with model.train().

6. Pass the training DataLoaders to MISTnet2 and record the loss values.

7. Back-propagate the loss and step through the optimizer to reparametrize MISTnet2.

8. Clear the optimizer gradients to refresh for the next epoch.

9. Set MISTnet2 to evaluation model with model.eval().

10. Pass the validation DataLoaders to MISTnet2 and record the loss.

11. Save out the model with the best validation loss performance.

12. Repeat steps 5 – 11 for 500 epochs.

13. Once training completes, load the model with the best validation performance.

14. Load the test datasets into additional DataLoaders.

15. Pass All DataLoaders through MISTnet2.

16. Save all outputs for further performance analysis.

**Objective Function (weighted MSE loss):**

1. Calculate the squared error across the predicted and ground-truth temperature volumes.

2. Scale the error at ground-truth regions between 300 and 600 Kelvin by 5.

3. Scale the error at ground-truth regions above 600 Kelvin by 10.

4. Return the mean of the scaled squared error.

## MISTnet2 MODEL ARCHITECTURE

Function custom_padding(3D volume, padding_amount):
    Pad the edges of the 3D volume in the shock-direction with chosen constant values the width of padding_amount.
    Pad the edges of the 3D volume periodically in the directions lateral to shock the width of padding_amount.
    Return the padded 3D volume.

Class padded_3D_Convolution:
    Initialize(input_channels, output_channels, kernel_size, padding_amount):
        Create a 3D convolutional layer with correlating kernel and channel sizes and no built-in padding.
        Store the padding amount.
    Forward(3D volume):
        Apply custom_padding to the 3D volume with the padding_amount.
        Apply a 3D convolution to the padded 3D volume.
        Return the convolution padded 3D volume.

Class Double_3D_Convolution:
    Initialize(input_channels, output_channels, residual_connection):
        Store channel sizes.
        Sequentially ordered sequence:
            Apply a Padded_3D_convolution.
            Apply a Group normalization.
            Apply a GELU activation.
            Apply a Padded_3D_convolution.
            Apply a Group normalization.
            If residual_connection is true, add the residual map here.
            Apply a GELU activation.
    Forward(3D volume):
        Pass the 3D volume through the sequential sequence.
        Return the processed 3D volume.

Class DownSample_block:
    Initialize(input_channels, output_channels, embedding_size):
        MaxPooling Sequence:
            Apply a 3D maxpool with kernel size = 2.
            Apply a Double_3D_Convolution.
            Apply a Double_3D_Convolution.
        Embedding layers (one for each embedding):
            Apply a SiLU activation.

Apply a Linear layer mapping embedding_size to output_channels.

Forward(microstructure_3D, lx, ly, lz, shock_strength, sim_type):
    Pass the microstructure_3D through the MaxPooling sequence.
    Pass lx, ly, lz, shock_strength, and sim_type through embedding layers.
    Reshape the embedding features to the max pooled microstructure_3D shape by repeating the outputs of embedding layers.
    Add the embedded features to the max pooled 3D volume
    Return the processed 3D volume

Class UpSample_block:
    Initialize(input_channels, output_channels, embedding_size):
        UpSampling:
            Apply a Trilinear upsample with scale factor = 2.
        Convolutional Sequence:
            Apply a Double_3D_Convolution.
            Apply a Double_3D_Convolution.
        Embedding layers (one for each embedding):
            Apply a SiLU activation.
            Apply a Linear layer mapping embedding_size to output_channels.

    Forward(microstructure_3D, skip_connection, lx, ly, lz, shock_strength, sim_type):
        Pass the microstructure_3D through the UpSampling.
        Concatenate the up-sampled 3D volume with the skip_connection.
        Pass the concatenation through the Convolutional Sequence.
        Pass lx, ly, lz, shock_strength, and sim_type through embedding layers.
        Reshape the embedding features to the upsampled microstructure_3D shape.
        Add the embedded features to the upsampled, concatenated, convoluted 3D volume.
        Return the processed microstructure_3D.

Class MISTnet2_Model:
    Initialize(input_channels=3, output_channels=1, embedding_size=256):
        Base_channel_size = 32.
        Encoder pathway:
            Input Double Convolution mapping input_channels to Base_channel_size.
            Down1: DownSample_block, map Base_channel_size to 2* Base_channel_size.
            Down2: DownSample_block, map 2*Base_channel_size to 4* Base_channel_size.
            Down3: DownSample_block, map 4*Base_channel_size to 4* Base_channel_size.
        Decoder pathway:
            Up1: UpSample_block, use Down3 output as skip connection,
                Mapping 8* Base_channel_size to 2*Base_channel_size.
            Up2: UpSample_block, use Down2 output as skip connection,
                Mapping 4* Base_channel_size to Base_channel_size.
            Up3: UpSample_block, use Down1 output as skip connection,
                Mapping 2* Base_channel_size to Base_channel_size.
        Final Output Layer:
            3D convolution with kernel_size = 1, mapping Base_channel_size to a single channel.
        Embedding processing layers (one for each embedding):
            Linear layer mapping original embedding size (1) to embedding_size.

    Forward(microstructure_3D, lx, ly, lz, shock_strength, sim_type):
        Pass embeddings through embedding processing layers, use relu activation.

Pass the microstructure_3D with processed embedding through the Encoder, save skip_connections.

Pass the encoded microstructure_3D, skip_connections, and processed embeddings through the Decoder.

Pass the Decoded microstructure_3D through the Final Output Layer

Return the Final Output microstructure_3D.

## 4. MISTnet2 Datasets:

The dataset used to train, validate, and test MISTnet2 consists of PBX, multi-pore, and single-pore RDX based systems that are shocked using a molecular dynamics framework at a particle velocity of 1.0, 1.5, or 2.0 km/s.

Schematic representations of all system are given below.

There are **7 PBX systems** shocked at a particle velocity of 1.0, 1.5, and 2.0 km/s using atomistic simulation. System 1, 3, 5, and 7 are used for training. System 4 and 6 are used for validation. System 2 is used for testing. Each system contains a top and bottom set.

System 1 (training):

## Bottom

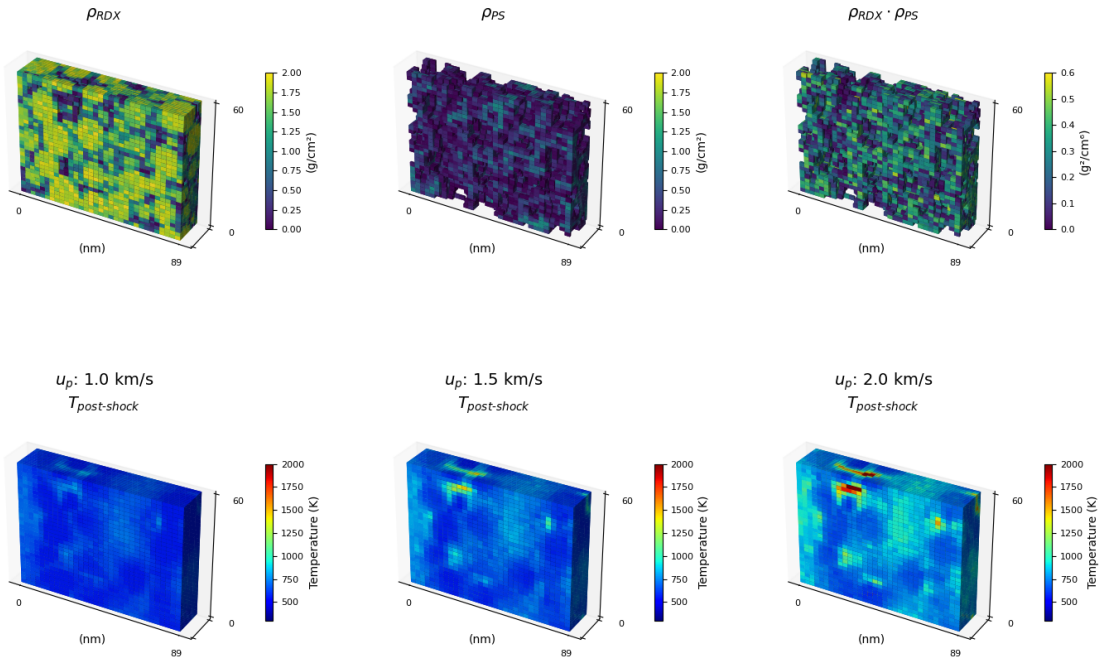

## Top

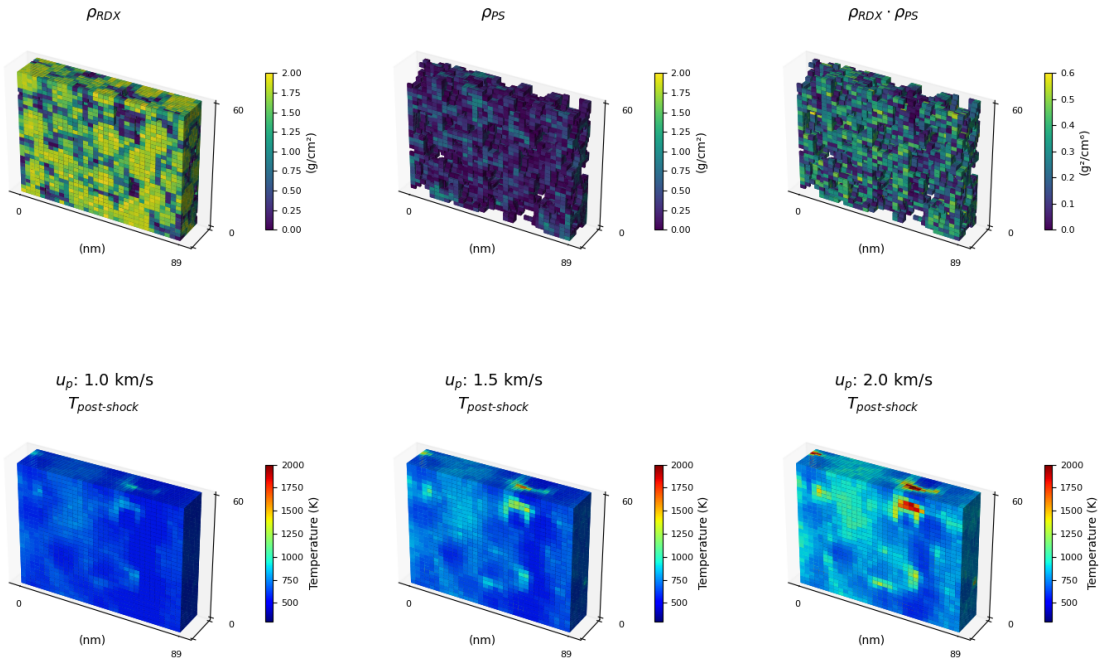

System 2 (testing):

## Bottom

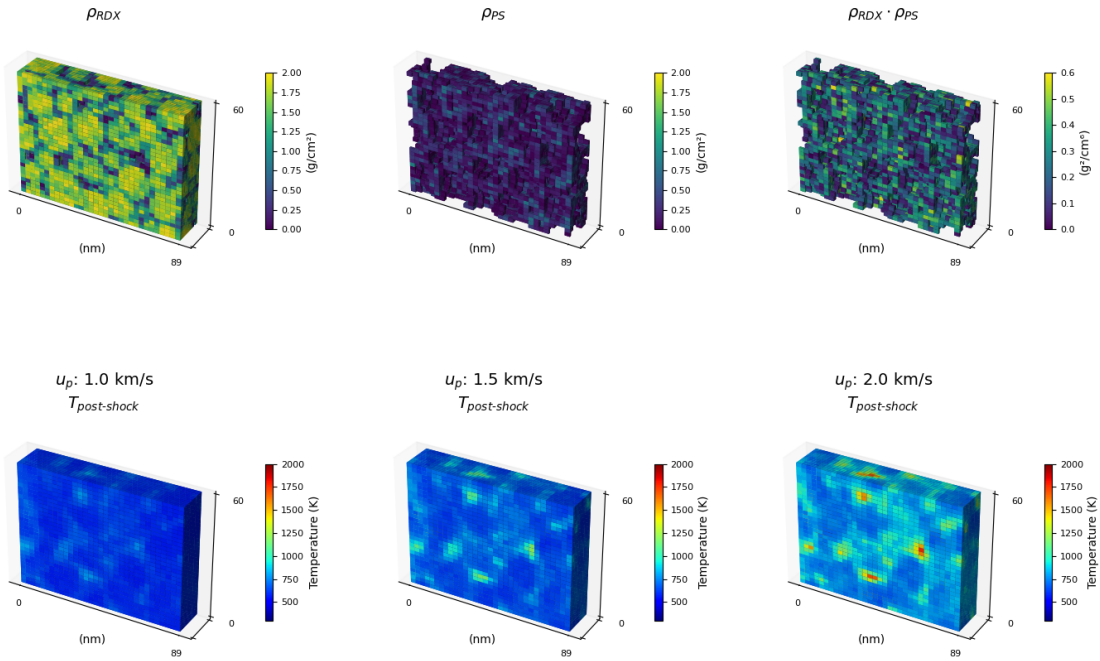

## Top

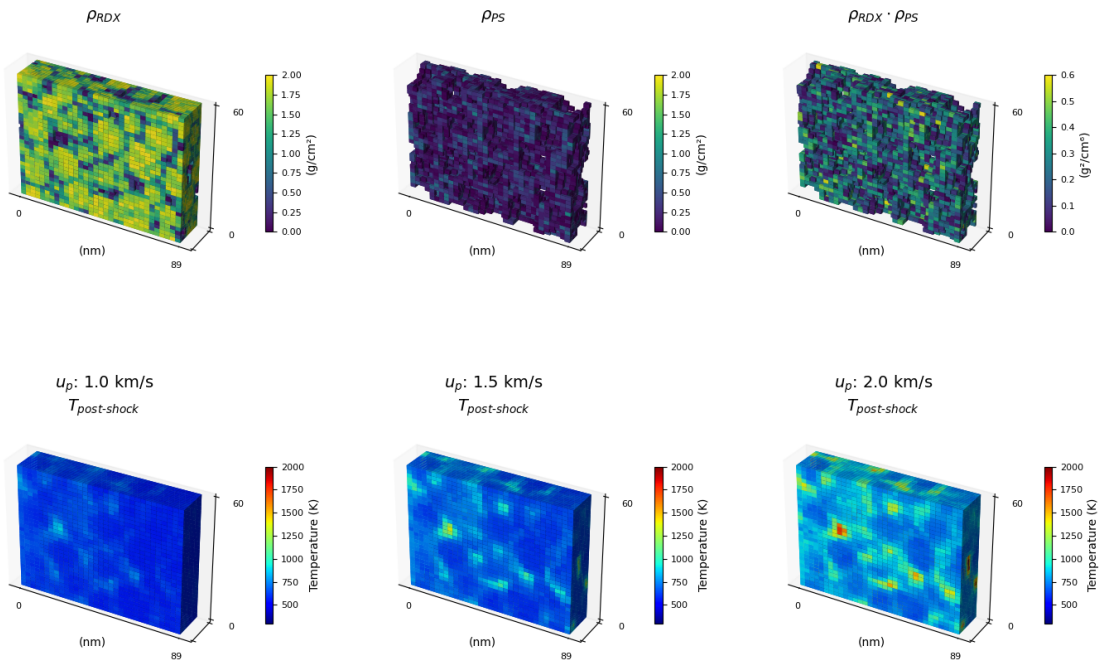

System 3 (training):

## Bottom

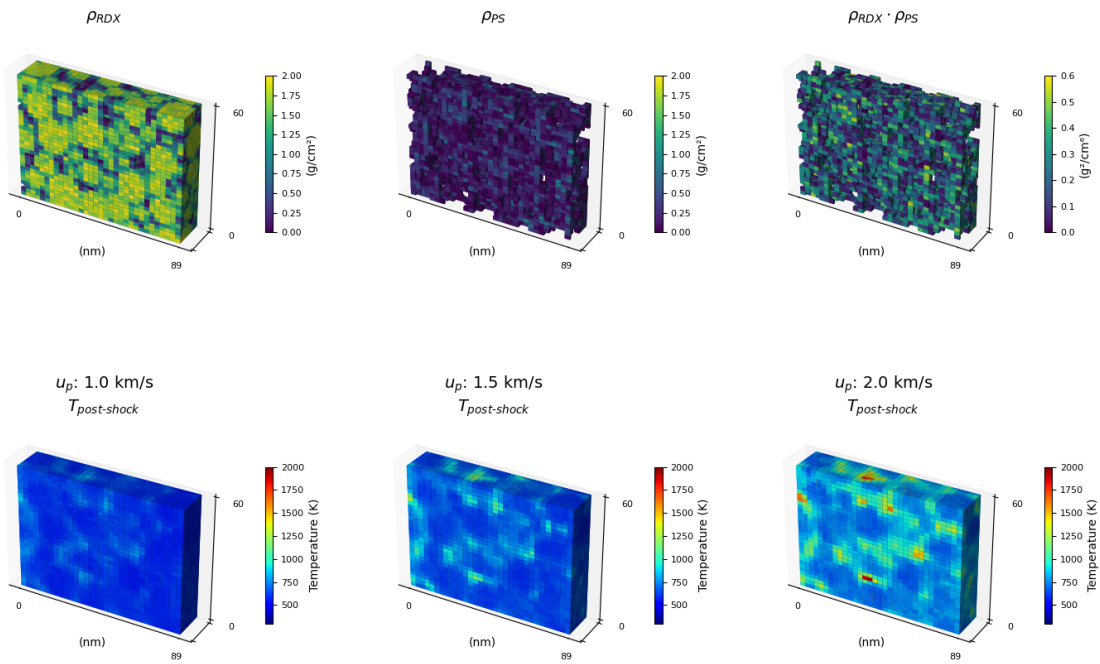

## Top

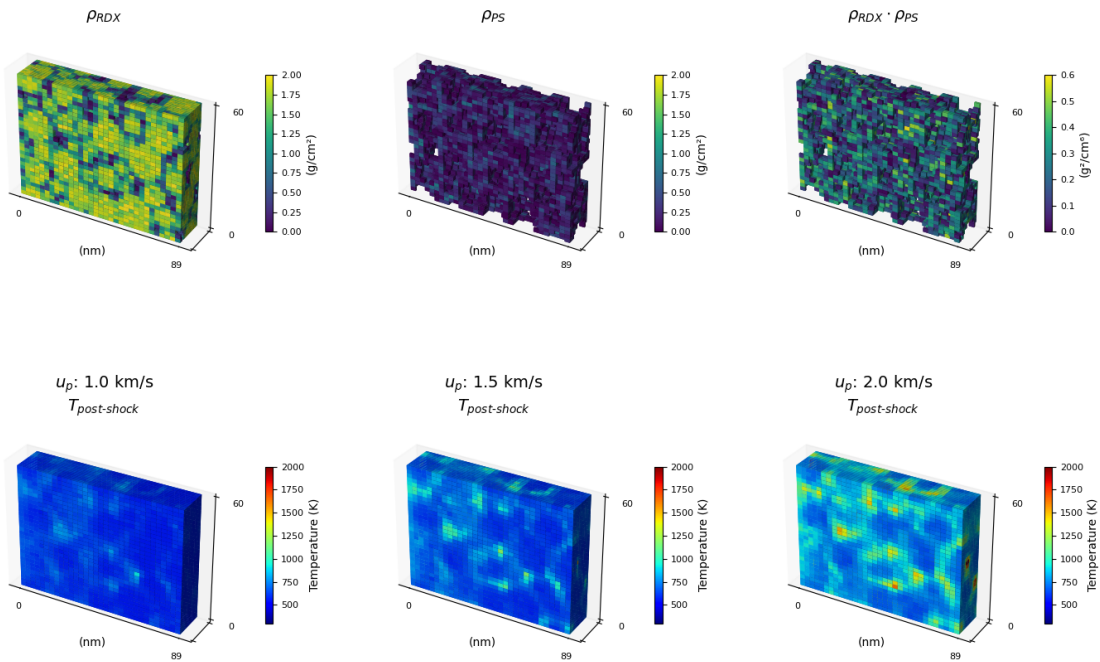

System 4 (validation):

## Bottom

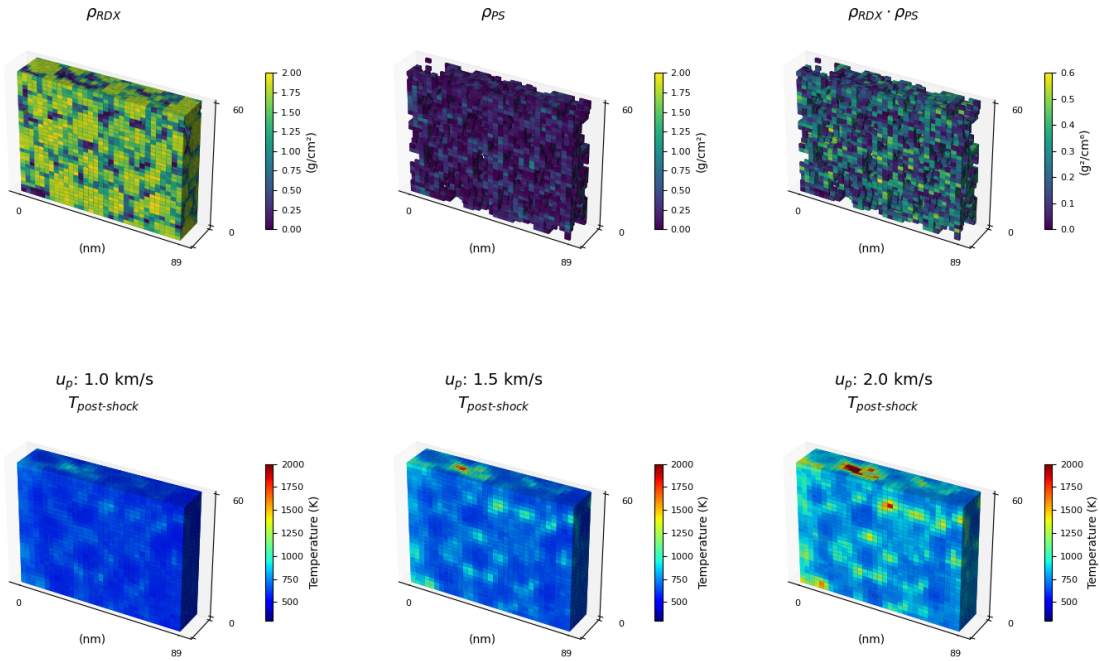

## Top

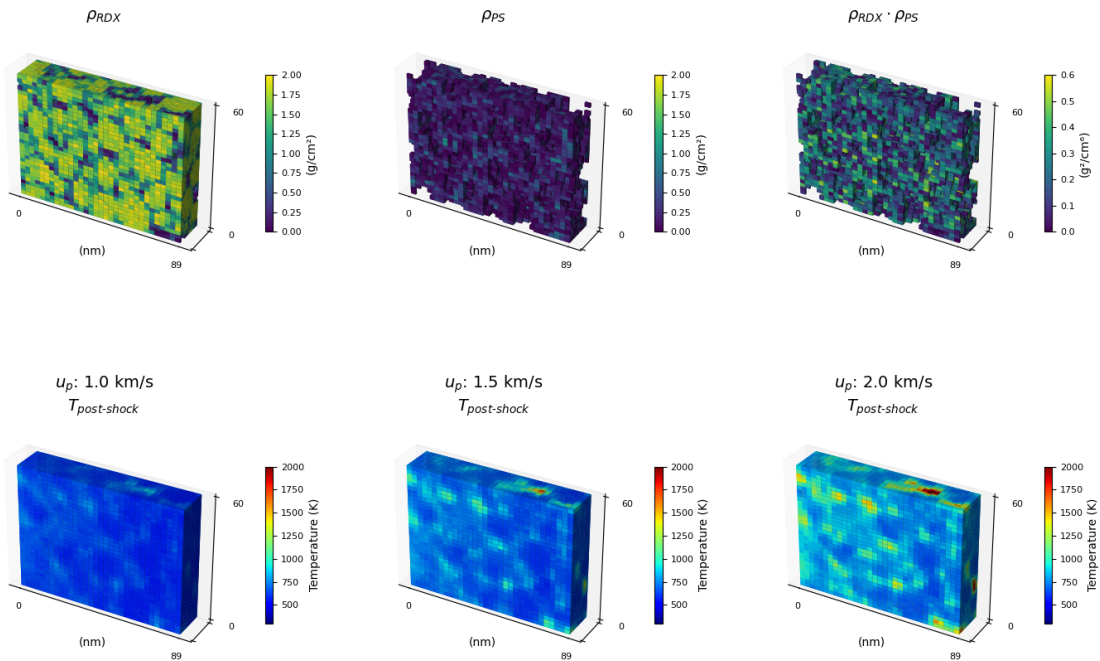

System 5 (training):

## Bottom

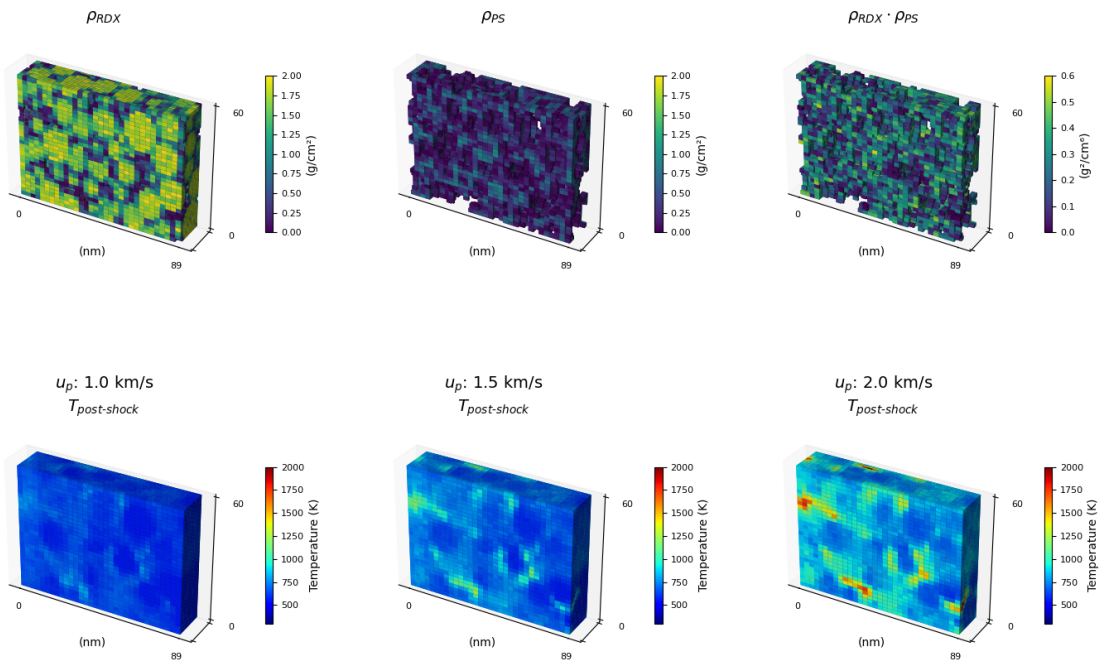

## Top

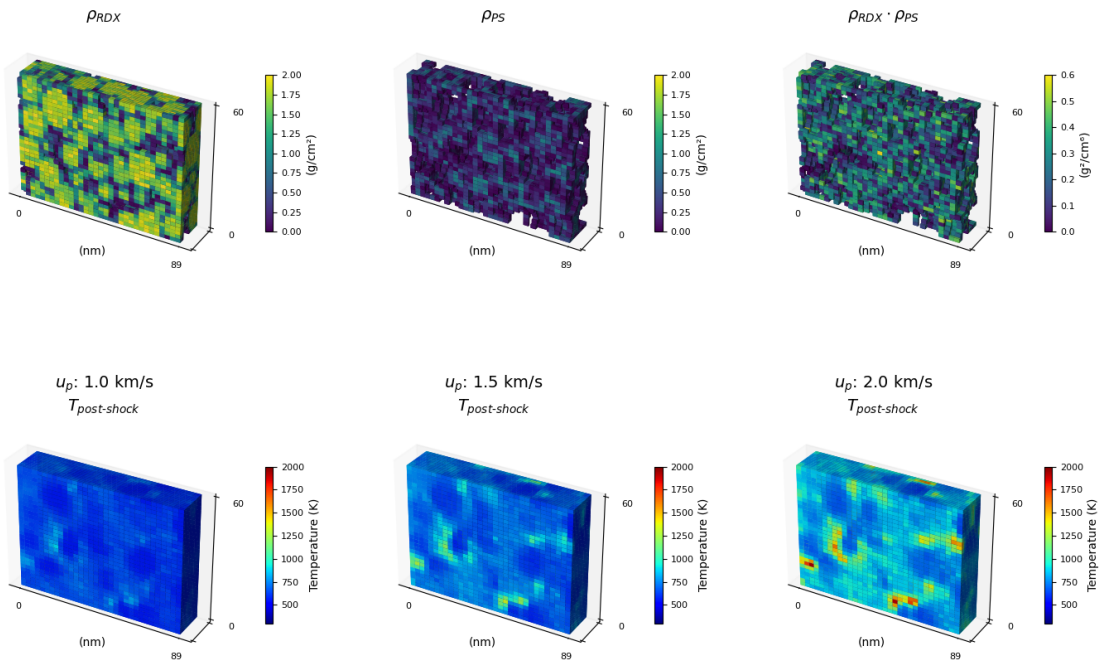

System 6 (validation):

## Bottom

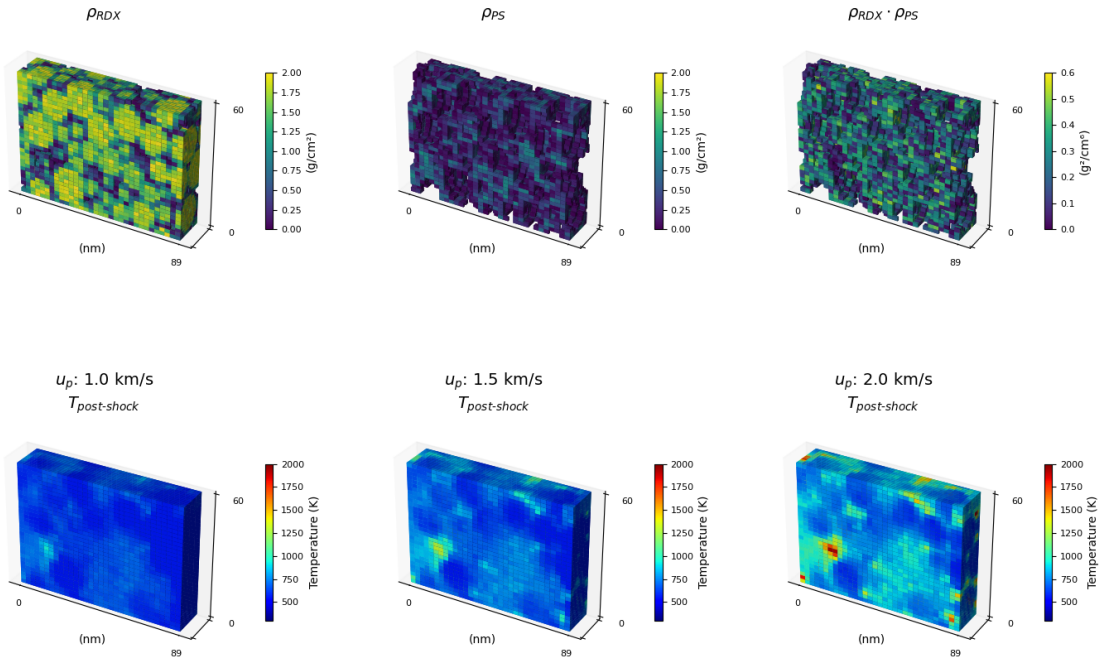

## Top

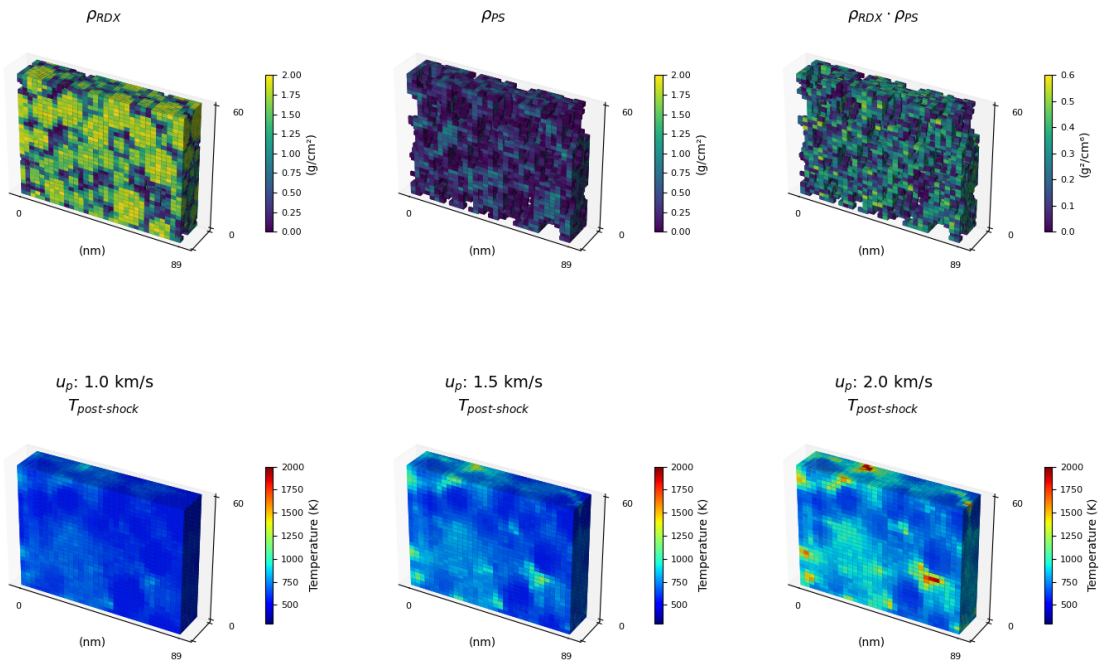

System 7 (training):

## Bottom

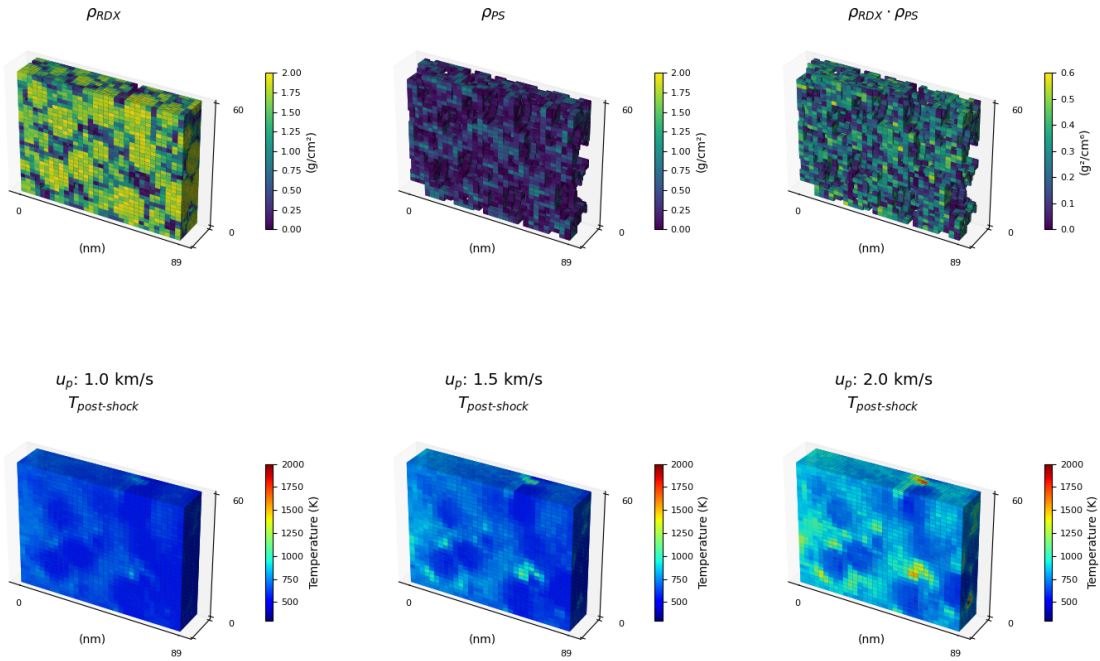

## Top

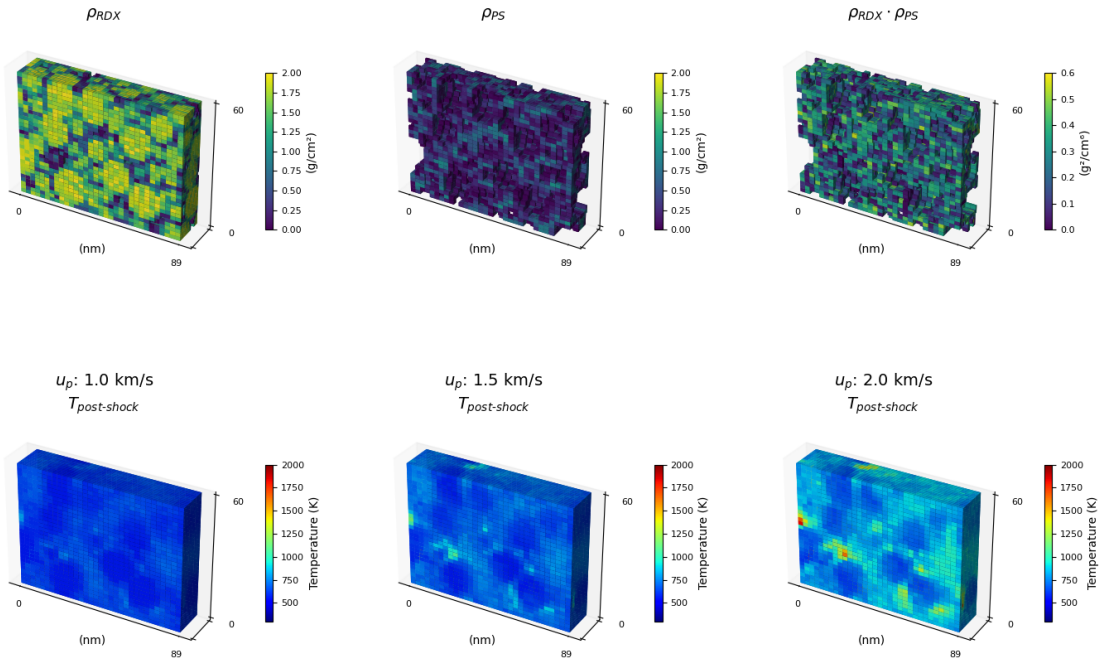

There are **10 multi-pore systems** shocked at a particle velocity of 1.0, 1.5, and 2.0 km/s using dissipative particle dynamic simulation. System 1, 2, 3, 4, 5, and 6 are used for training. System 9 and 10 are used for validation. System 7 and 8 are used for testing.

System 1 (training):

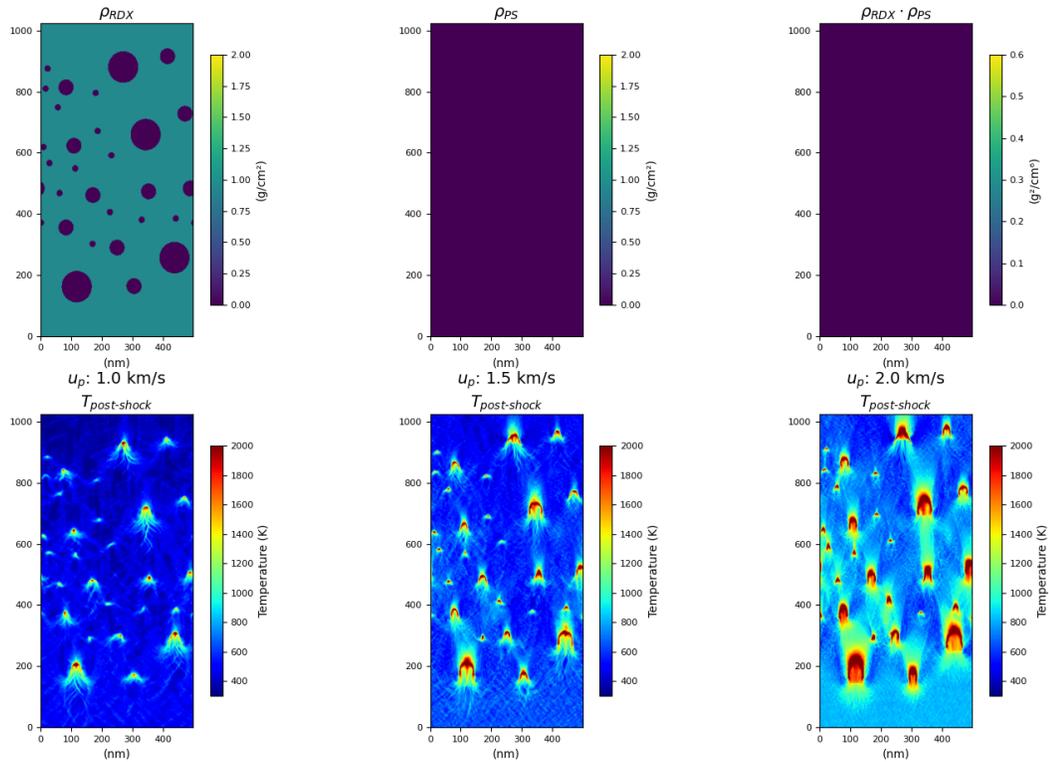

System 2 (training):

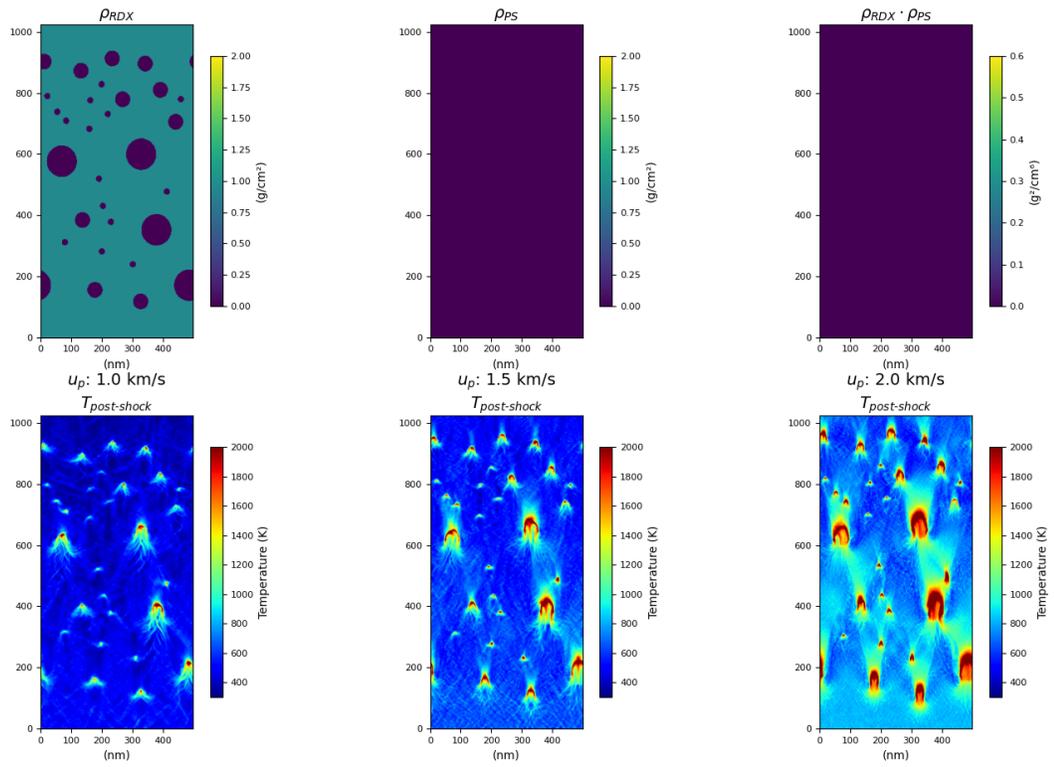

System 3 (training):

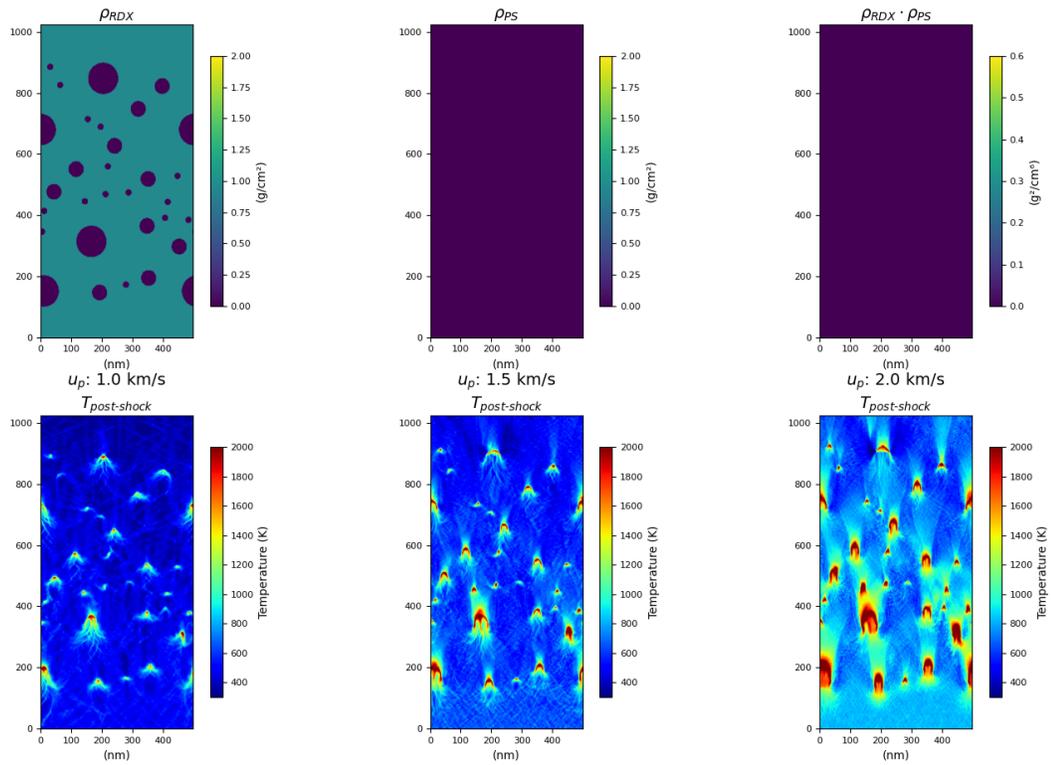

System 4 (training):

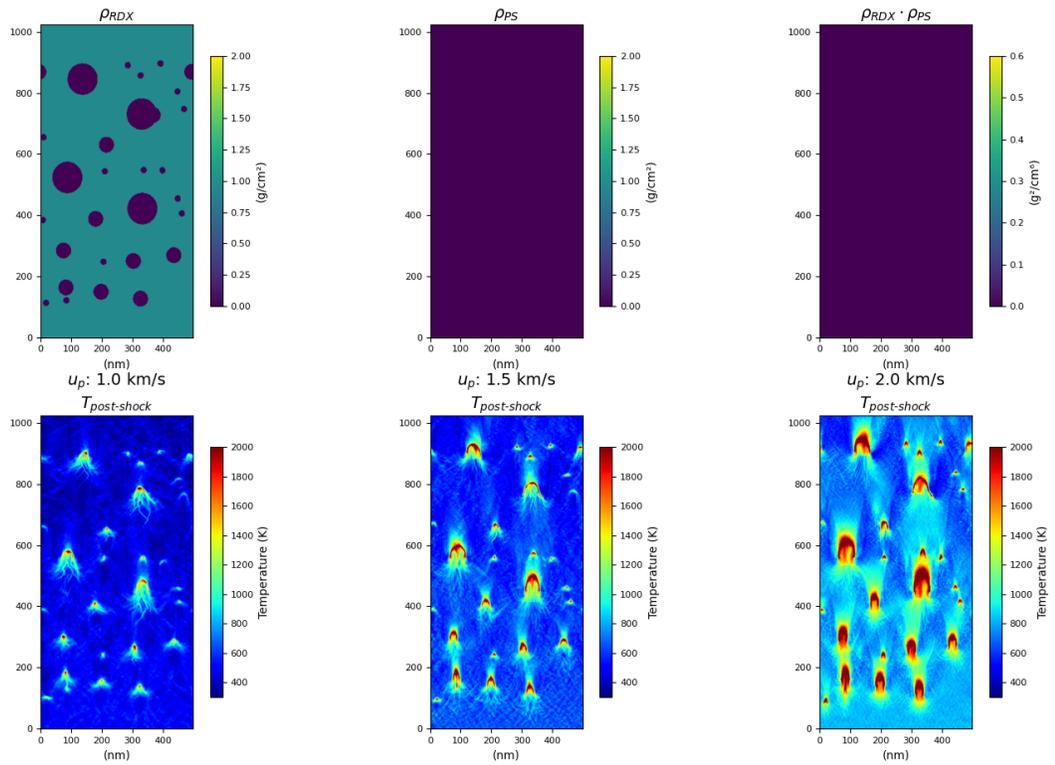

System 5 (training):

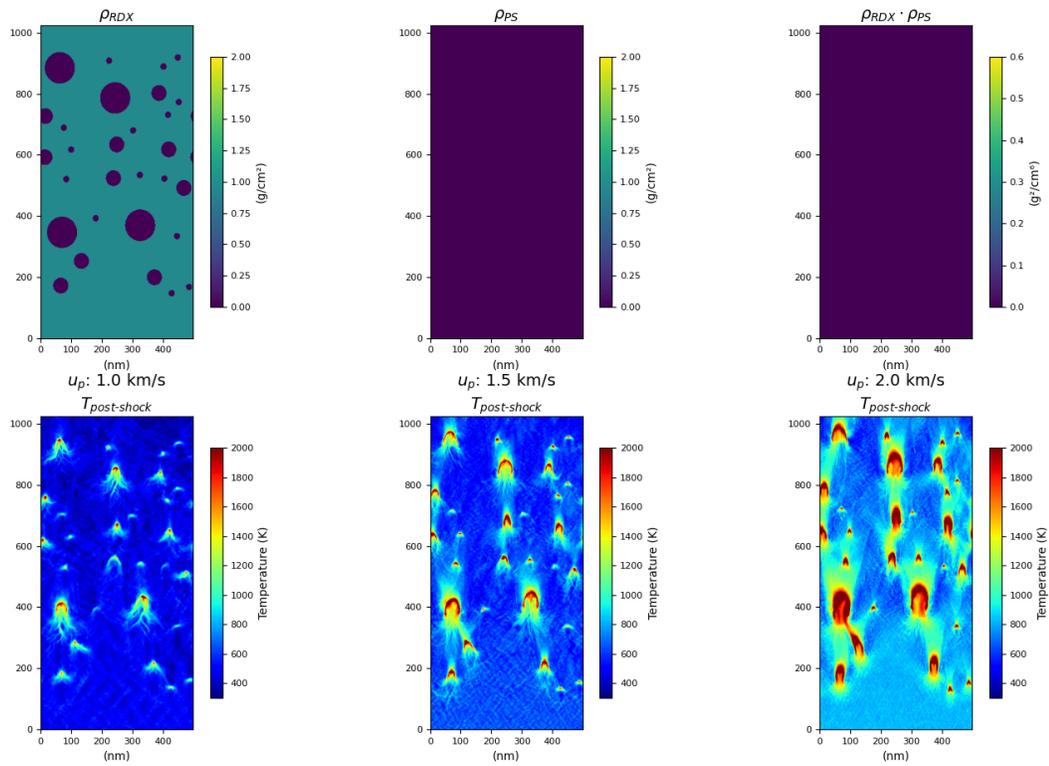

System 6 (training):

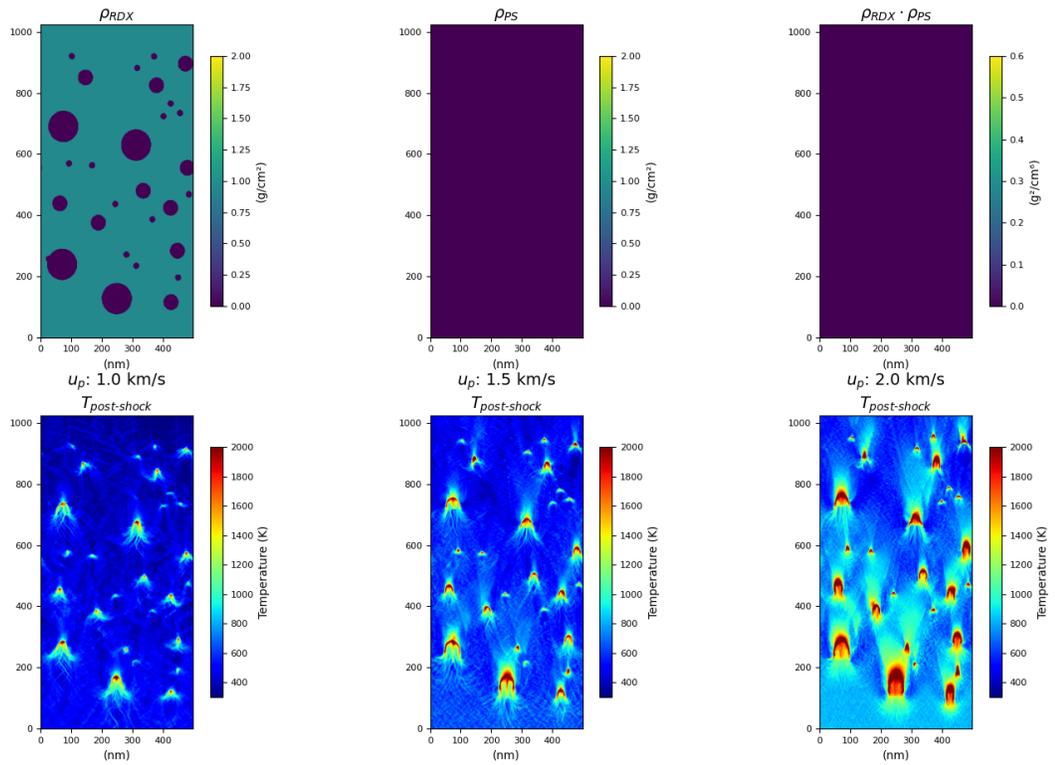

System 7 (testing):

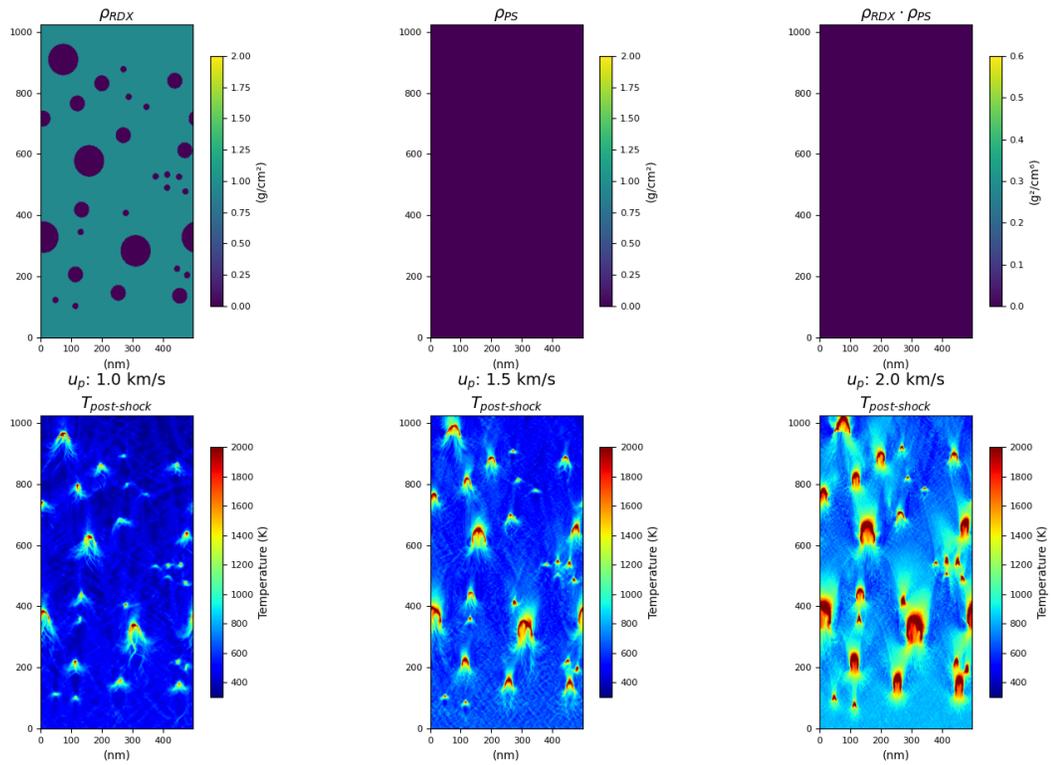

System 8 (testing):

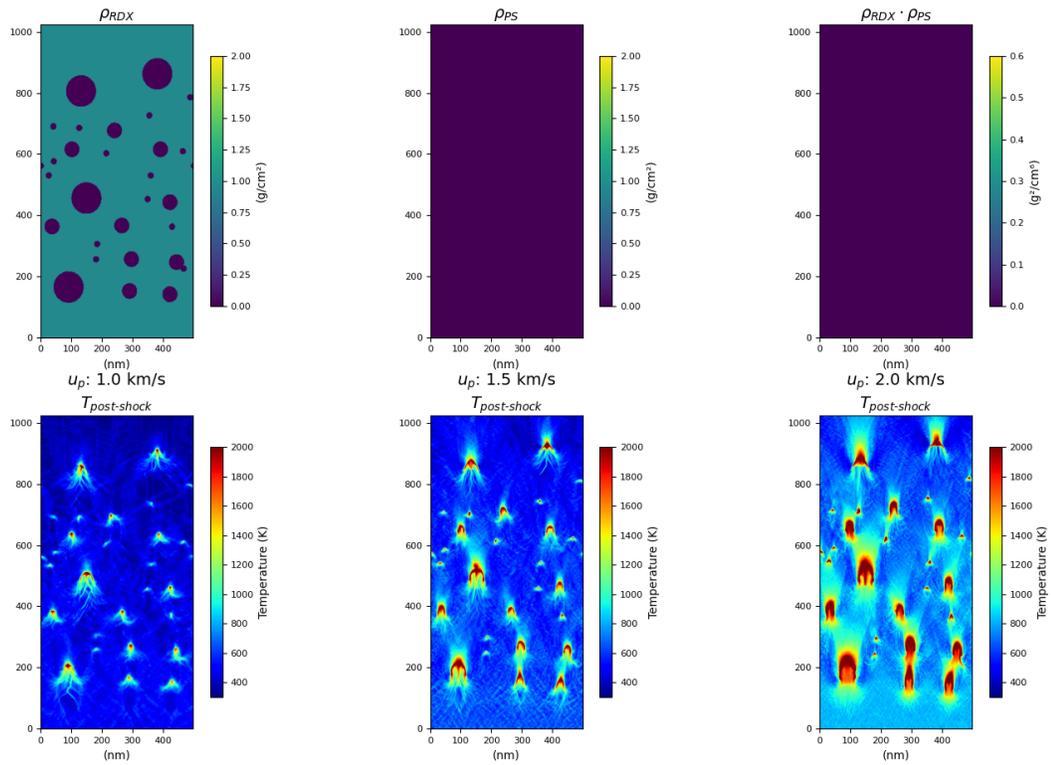

System 9 (validation):

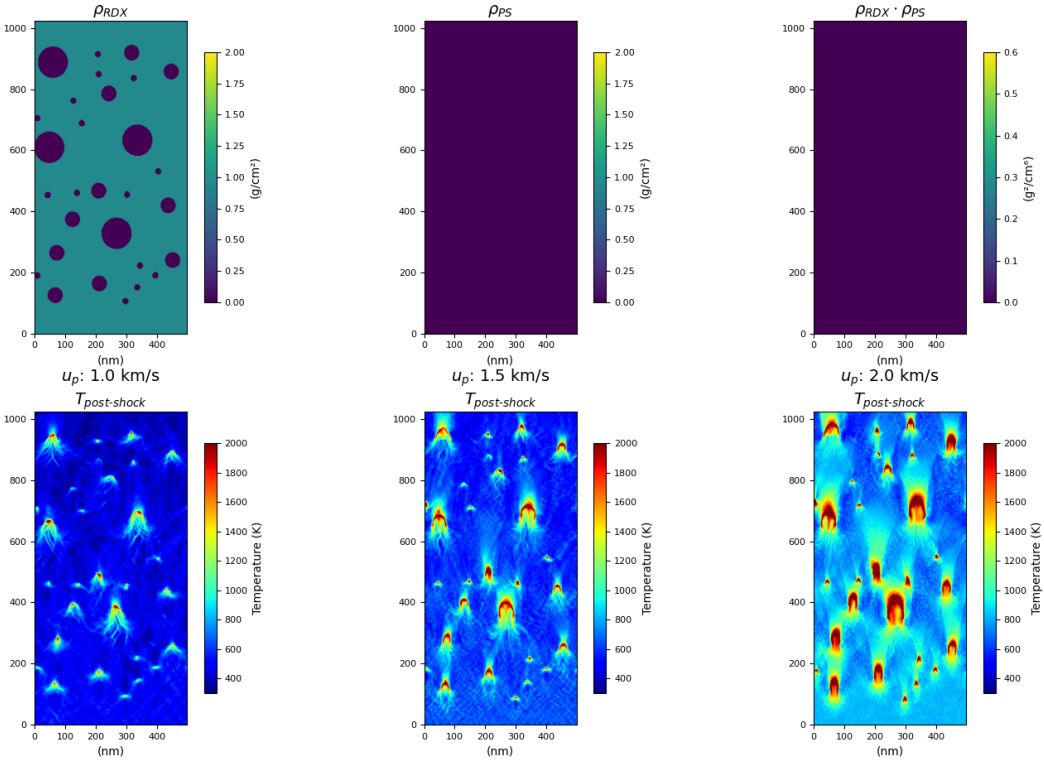

System 10 (validation):

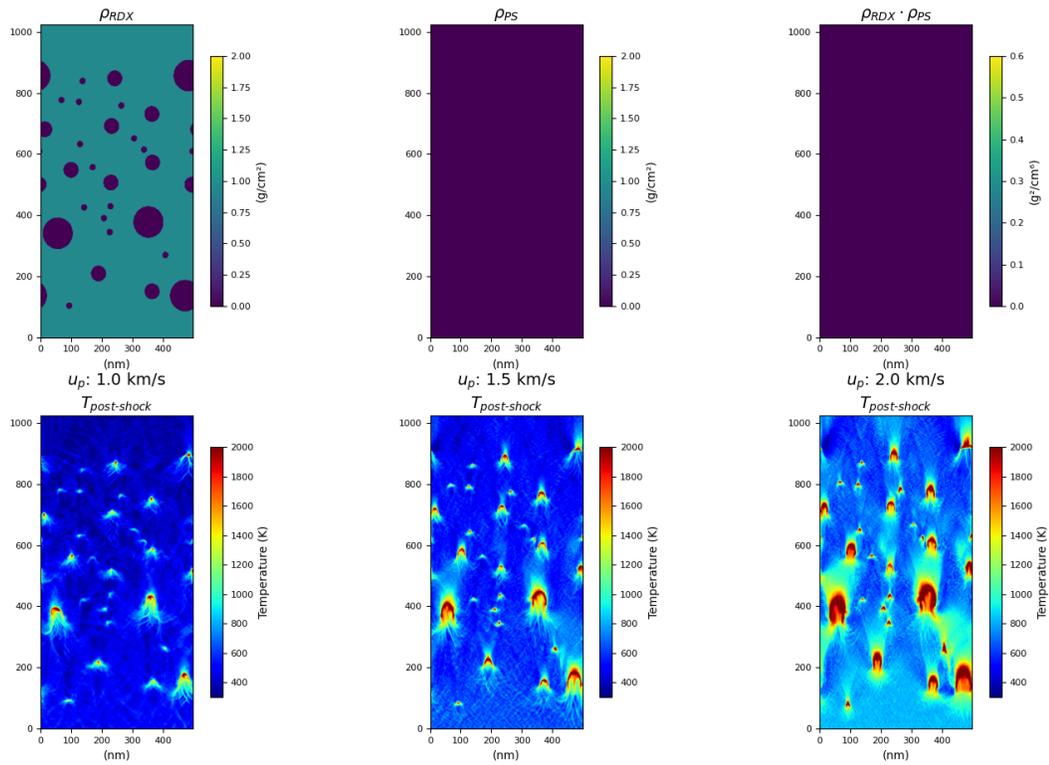

There are **3 single-pore systems** shocked at a particle velocity of 1.0, 1.5, and 2.0 km/s using dissipative particle dynamic simulation. All are used for training.

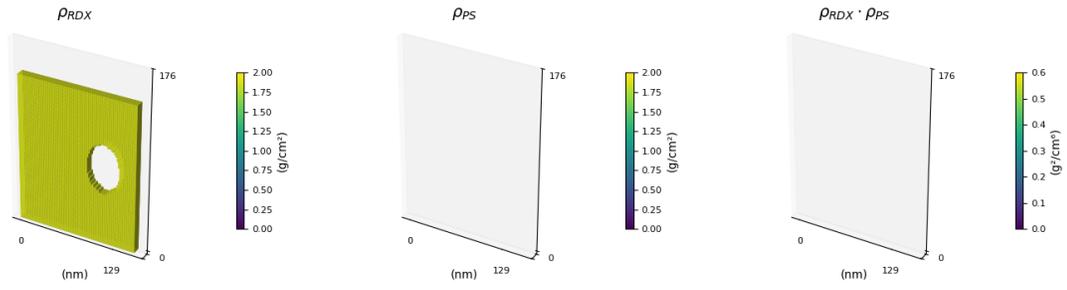

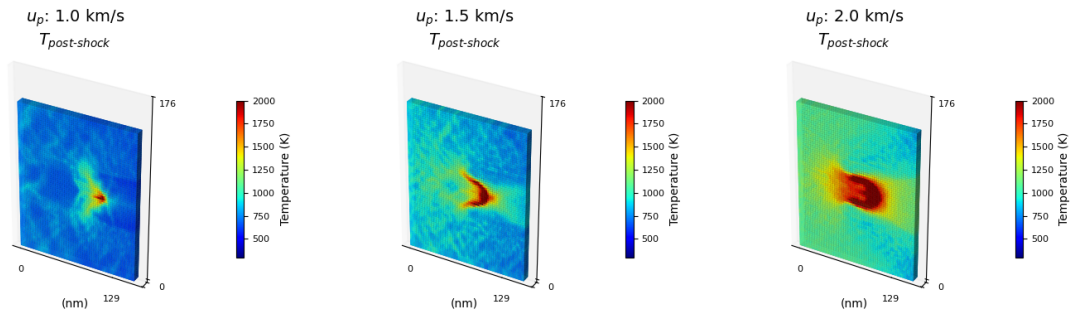

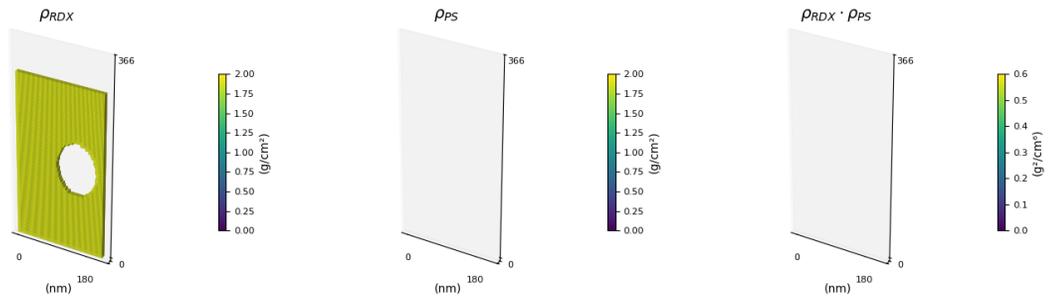

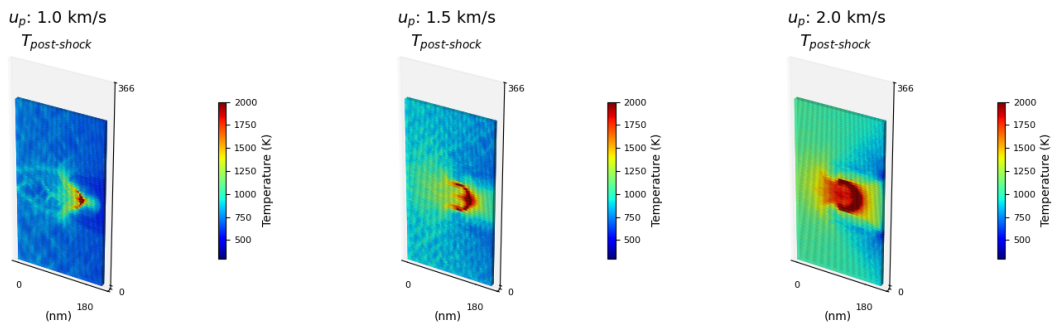

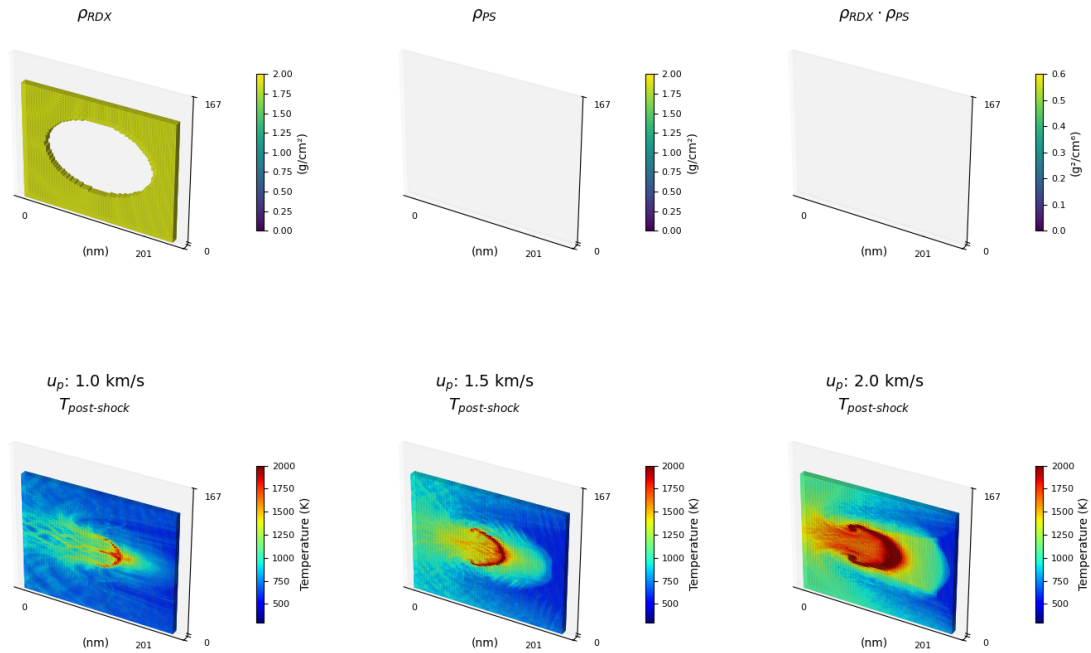

There are **3 single-pore systems** shocked at a particle velocity of 1.0 and 2.0 km/s using atomistic simulation. All are used for training.

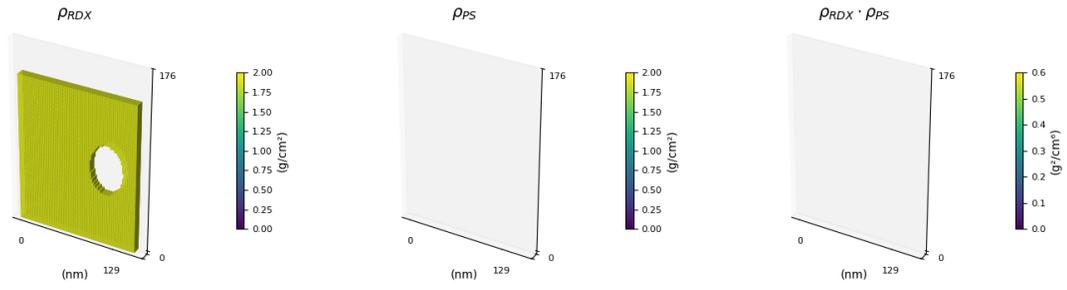

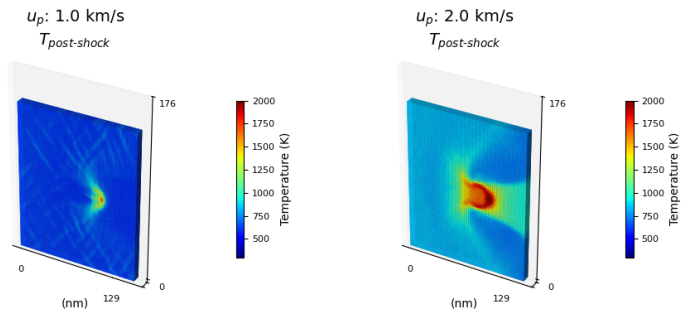

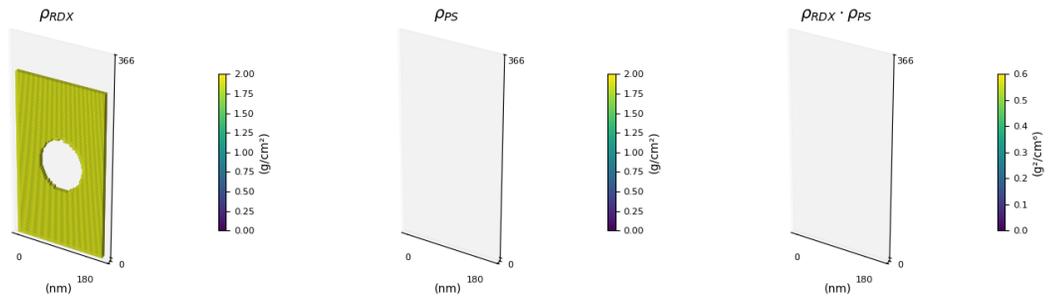

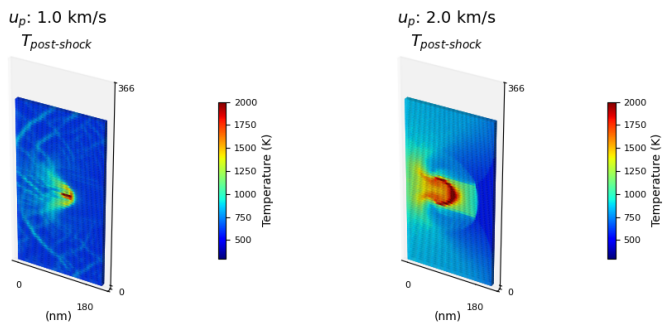

$\rho_{RDX}$

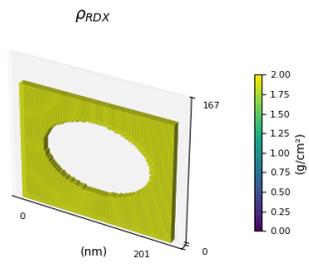

$\rho_{PS}$

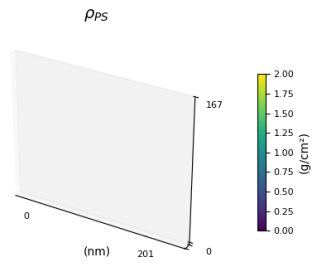

$\rho_{RDX} \cdot \rho_{PS}$

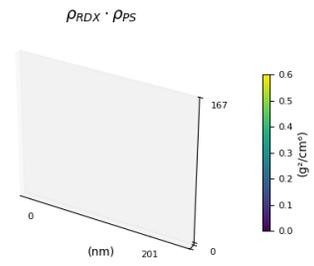

$u_p$: 1.0 km/s
$T_{post\text{-}shock}$

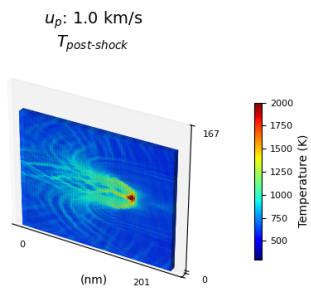

$u_p$: 2.0 km/s
$T_{post\text{-}shock}$

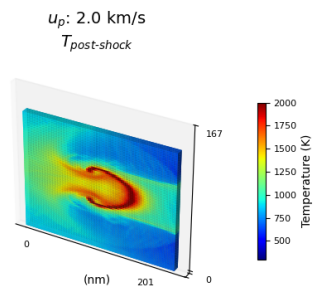